\documentclass[12pt]{article}
\usepackage{epsfig,amssymb}
\usepackage{latexsym}

\newcommand{\bq}{\begin{equation}}
\newcommand{\eq}{\end{equation}}
\newcommand{\bqa}{\begin{eqnarray}}
\newcommand{\eqa}{\end{eqnarray}}
\newcommand{\ben}{\begin{enumerate}}
\newcommand{\een}{\end{enumerate}}
\newcommand{\bc}{\begin{center}}
\newcommand{\ec}{\end{center}}
\newcommand{\bqb}{\begin{eqnarray*}}
\newcommand{\eqb}{\end{eqnarray*}}

\def\pr#1#2#3{Phys. Rev. ${\bf{#1}}$, #2 (#3)}
\def\prl#1#2#3{ Phys. Rev. Lett. ${\bf{#1}}$, #2 (#3)}

\def\prep#1#2#3{ Phys. Rep. ${\bf{#1}}$, #2 (#3)}

\def\np#1#2#3{ Nucl. Phys. ${\bf{#1}}$, #2 (#3)}
\def\zp#1#2#3{ Z. f. Phys. ${\bf{#1}}$, #2 (#3)}

\def\epj#1#2#3{ Eur. Phys. J. ${\bf{#1}}$, #2 (#3)}
\def\ijmp#1#2#3{Int. J. Mod. Phys. ${\bf{#1}}$, #2 (#3)}

\def\fortp#1#2#3{Fortsch. Phys. ${\bf{#1}}$, #2 (#3)}

\def\aop#1#2#3{Annals of Phys. ${\bf{#1}}$, #2 (#3)}

\def\polon#1#2#3{Acta Phys. Polon. ${\bf{#1}}$, #2 (#3)}
\def\lnc#1#2#3{Lett.Nuov.Cim. ${\bf{#1}}$, #2 (#3)}
\begin{document}
\pagenumbering{arabic}
\thispagestyle{empty}
\def\thefootnote{\fnsymbol{footnote}}
\setcounter{footnote}{1}

%\vspace{2cm}
\begin{flushright}
November 1, 2013\\
arXiv: 1309.3177[hep-ph]\\
 \end{flushright}

\vspace{2cm}

\begin{center}
{\Large {\bf A  supersimple analysis of $e^-e^+\to  W^- W^+$\\
 at high energy}}.\\
 \vspace{1cm}
%-----------------------------------------------------------------
{\large G.J. Gounaris$^a$ and F.M. Renard$^b$}\\
\vspace{0.2cm}
$^a$Department of Theoretical Physics, Aristotle
University of Thessaloniki,\\
Gr-54124, Thessaloniki, Greece.\\
\vspace{0.2cm}
$^b$Laboratoire Univers et Particules de Montpellier,
UMR 5299\\
Universit\'{e} Montpellier II, Place Eug\`{e}ne Bataillon CC072\\
 F-34095 Montpellier Cedex 5.\\
\end{center}

\vspace*{1.cm}
\begin{center}
{\bf Abstract}
\end{center}

Studying  $e^-e^+\to  W^- W^+$ at the 1loop electroweak (EW) order,
we derive very accurate and simple expressions for the
four Helicity Conserving (HC) amplitudes, which  dominate  this process
at high energies. The calculations are done in both, the SM and MSSM frameworks.
Such  expressions, called supersimple (sim),
nicely  emphasize the  dynamical contents of each framework.
Numerical illustrations are presented, which show the accuracy of this  description,
and how it can be  used for identifying possible additional new physics contributions;
like e.g. Anomalous Gauge Couplings (AGC) or a new  $Z'$ vector boson exchange.
The procedure is  useful even if only SM is visible at the future Linear Collider energies.

\vspace{0.5cm}

\def\thefootnote{\arabic{footnote}}
\setcounter{footnote}{0}
\clearpage

\section{ Introduction}

  The process $e^-e^+\to W^-W^+$ has been studied theoretically
  and experimentally since a long time, as it provides sensitive
  tests of the gauge structure of the
  electroweak interactions \cite{Boehm, Beenakker,Hahn, Denner},  and  checks the possible
  presence of non standard new physics (NP) contributions.
  A detail history of the subject and a list of references may be seen   in \cite{Denner}.

  First experimental studies of this process have been done at LEP2 \cite{LEP2WW}.
  No signal of departures from SM has been found, but the accuracy is not sufficient
  to eliminate the possibility of  NP effects at a  high scale.

  LHC studies  involving production of  W pairs also exist; but their detailed studies
  require a difficult event analysis, because of various sources of background
  \cite{LHCWW}.

  Future high energy  $e^-e^+$ colliders are therefore deeply desired,
  in order to provide fruitful information about this subject \cite{ILC,CLIC}.

  From  the theoretical side, the present situation of an 1loop electroweak  (EW)
  order  analysis,   aiming e.g. at  searching
  for any non standard effects, is quite   complex.
  This is already true at  the SM level, and if one  includes
  SM extensions, like e.g. SUSY,   the sensitivity to any benchmark
  choice has to be considered. Particularly for
  amplitudes involving longitudinal $W$'s, the numerical situation is more difficult,
  because of the huge cancelations taking place. In both the SM and SUSY  cases,
  very lengthy numerical codes are required to describe
  the complete 1loop EW contribution; see e.g.   \cite{Hahn, Denner}.

  The aim of the present paper is to call attention to the fact that at high energies,
  the 1loop electroweak (EW) corrections to the helicity amplitudes for $e^-e^+\to W^-W^+$,
  acquire very simple forms, in  both, the SM and MSSM cases. To establish them we have
  done a complete calculation of the  1loop diagrams and then taken the high energy,
  fixed angle, limit using \cite{asPV}. The soft photon bremsstrahlung can then
  be added as usual \cite{Boehm, Beenakker,Hahn, Denner}.

  Our procedure is the same as the one   used previously for other 2-to-2 processes,
  leading to the "supersimple" (sim) 1loop EW expressions  for the dominant high energy
  helicity conserving (HC)  amplitudes; the   helicity violating (HV)
  ones  are quickly vanishing\footnote{The notations HC and HV are fully
   defined in the next section.}
   \cite{super, ttbar}. We find very simple and quite accurate
   expressions for the high energy HC amplitudes,  in both the SM and MSSM frameworks, which
  nicely  show  their relevant  dynamical contents.

  The use of this  description, which clearly indicates the relevant physical parameters,
  should very much simplify  the analysis of the experimental results. Particularly because,
      its  accuracy turns out to  be
    sufficient for distinguishing   1loop SM (or MSSM) effects,
  from e.g. various  types of additional New Physics contributions, like AGC couplings
  or $Z'$ exchange; see for example \cite{Andreev}.

  The content of the paper is the following. In Section 2 we present
  the various properties of the high energy $e^-e^+\to W^-W^+$ amplitudes,
  with  special attention to their helicity conservation
  (HCns) property \cite{heli1, heli2}. The explicit  supersimple  expressions
  are discussed in the later part of Section 2 and in Appendix A.
  In Section 3 we present  the   energy and angular dependencies
  of the  cross sections, for  polarized and unpolarized electron beams,
  in either SM or MSSM. And subsequently, we compare these SM or MSSM contributions
  to  those due to  anomalous gauge couplings (AGC) or  $Z'$ effects; both    given
  in Appendix B. We find  that the accuracy of the supersimple
  expressions is sufficient for distinguishing these various types
  of  contributions. Thus, they may be used instead of the complete 1loop results.
  The conclusions summarize these results.\\

\section{Supersimplicity in  $e^-e^+\to W^-W^+$}

The process studied, to the 1loop Electroweak (EW) order, is
\bq
e^-_\lambda (l) ~ e^+_{\lambda'} (l') \to W^-_\mu (p)~ W^+_{\mu'}(p')~~, \label{process}
\eq
where $(\lambda, \lambda')$ denote the helicities of the incoming $(e^-, e^+)$ states,
and $(\mu, \mu')$ the helicities of the outgoing $(W^-, W^+)$. The corresponding
momenta are denoted as $(l,l',p,p')$. Kinematics are defined through
\bqa
 && s=(l+l')^2=(p+p')^2 ~~,~~ t=(l-p)^2=(l'-p')^2 ~~,~~ \nonumber \\
 && p_W=\sqrt{{s\over4}-m^2_W}~~,~~ \beta_W=\sqrt{1-\frac{4 m_W^2}{s}} ~~, \label{kinematics}
\eqa
where $p_W, \beta_W$ denote respectively the $W^\mp$ three-momentum and velocity in the
$W^-W^+$-rest frame. Finally,   the angle between the incoming $e^-$
momentum $l$ and the outgoing $W^-$ momentum $p$, in the  center of mass frame,
is denoted as $\theta$.

Due to the smallness of  the electron mass,
non-negligible amplitudes at high energies only appear for $\lambda=-\lambda'=\mp 1/2$.
The helicity amplitudes for
this process are therefore determined by three helicity indices and denoted
as $F_{\lambda,\mu,\mu'}(\theta)$, where $(e^-, W^-)$
are treated as particles No.1,  and $(e^+,W^+)$ as particles No.2, in the standard
Jacob-Wick notation \cite{JW}.

Assuming CP invariance, we obtain the constraint
\bq
F_{\lambda, \mu,\mu'}(\theta)=F_{\lambda, -\mu',-\mu}(\theta) ~~, \label{CP-cons}
\eq
which means that the process  is described by just 12 independent helicity
amplitudes.

At high energy, the helicity conservation (HCns) rule  implies
that only the amplitudes satisfying
\bq
\lambda +\lambda'=0= \mu + \mu' ~~, \label{heli-cons}
\eq
can dominate  \cite{heli1,heli2}.  These  are the
 helicity conserving  (HC) amplitudes, which explicitly  are
\bq
  F_{\mp-+}~~,~~ F_{\mp+-}~~,~~ F_{\mp 00} ~~.~~ \label{6HC-amp-list}
\eq
The purely left-handed $W$ couplings though,  forces the HC amplitudes
\bq
F_{++-}~~,~~F_{+-+}~~, \label{R-HC-amp-list}
\eq
to vanish at Born-level and be very small at 1loop.
Thus, only four  leading HC helicity amplitudes remain at high energy,
namely
\bq
F_{--+}~~,~~ F_{-+-}~~,~~ F_{\pm00}~~.~~ \label{4HC-amp-list}
\eq
The remaining amplitudes, which violate  (\ref{heli-cons}), are termed as
helicity violating (HV) ones. Explicitly these are
\bq
F_{-0+}~~,~~F_{---}~~,~~F_{-+0}~~,~~F_{+0+}~~,~~F_{+--}~~,~~F_{++0}~~,~~ \label{HV-amp-list}
\eq
 and are expected to be suppressed  like $m_W/\sqrt{s}$ ~ or $m^2_W/ s$, at
 high energy.\\

\subsection{Born contribution to the helicity  amplitudes}

   We next turn to the Born contribution to the HC  and HV amplitudes in (\ref{4HC-amp-list})
   and  (\ref{HV-amp-list}) respectively. The relevant diagrams involve
   neutrino exchange in the t-channel and photon+Z exchange in the s-channel. The resulting
   amplitudes    satisfy the   HCns constraints  \cite{heli1, heli2}.
 In the usual Jacob and Wick convention \cite{JW}, their  exact expressions are:\\

\noindent
{\bf Transverse-Transverse (TT)  amplitudes ($\mu,\mu'=\pm1$)}

Using (\ref{kinematics}), we find
\bqa
F^{\rm Born}_{\lambda \mu \mu' }&=& {se^2\sin\theta\over16ts^2_W}
\delta_{\lambda,-}
\left \{ \mu+\mu'+\beta_W(1+\mu\mu')-2\mu(1+\mu'\cos\theta) \right \}
\nonumber\\
&&+{se^2\over4}\left [{Q_e\over s}+{a_{eL}\delta_{\lambda,-}+a_{eR}\delta_{\lambda,+}\over
2s^2_W(s-m^2_Z)}\right ](1+\mu\mu')(2\lambda)\beta_W\sin\theta ~~,
\label{FBorn-TT}
\eqa
with
\bq
Q_e=-1~~,~~a_{eL}=1-2s^2_W~~,~~a_{eR}=2s^2_W~~, \label{e-couplings}
\eq
determining the electron charge, and the $Z$ left- and right-couplings.
Because of the purely left-handed $W$ coupling,  Eqs.(\ref{FBorn-TT}) leads to
\bq
F^{\rm Born}_{+{1\over2},\mu, -\mu}=0 ~~, \label{R-HC-TT-Born}
\eq
as already said  just after (\ref{R-HC-amp-list}).
In addition, (\ref{FBorn-TT}) leads  at high energy to
\bq
F^{\rm Born}_{\lambda\mu\mu} \to 0  ~~, \label{Born-asym-TT-HV}
\eq
in agreement with HCns  \cite{heli1, heli2}, and
\bq
F^{\rm Born}_{-{1\over2}\mu-\mu}\to - {e^2 \sin\theta (\mu-\cos\theta)
\over 4s^2_W (\cos\theta-1)}~~. \label{Born-asym-TT-HC}
\eq
This confirms that the first two HC Born amplitudes in (\ref{4HC-amp-list}),
go to constants, asymptotically.\\

\noindent
{\bf Transverse-Longitudinal (TL) and Longitudinal-Transverse (LT)\\
amplitudes ($\mu=\pm1,\mu'=0$,  $\mu=0,\mu'=\pm1$)}

Using again  (\ref{kinematics}), we obtain
\bqa
F^{\rm Born}_{\lambda\mu 0}&=& {s\sqrt{s}e^2\over8\sqrt{2}m_Wts^2_W}\delta_{\lambda,-}
\left \{ (\beta_W-\cos\theta)(1-\mu\cos\theta)- {2m^2_W\over s}(\mu-\cos\theta)\right \}
\nonumber\\
&&-{s\sqrt{s}e^2\over2\sqrt{2}m_W}\left [{Q_e\over s}+{a_{eL}\delta_{\lambda,-}
+a_{eR}\delta_{\lambda,+}\over
2s^2_W(s-m^2_Z)} \right ]\beta_W(1+2\lambda\mu\cos\theta)~~, \label{FBorn-TL} \\
F^{\rm Born}_{\lambda 0 \mu'}&=& {s\sqrt{s}e^2\over8\sqrt{2}m_Wts^2_W}\delta_{\lambda,-}
\left \{(\beta_W-\cos\theta)(1+\mu'\cos\theta)- {2m^2_W\over s}(\mu'+\cos\theta) \right \}
\nonumber\\
&&-{s\sqrt{s}e^2\over2\sqrt{2}m_W}
\left [{Q_e\over s}+{a_{eL}\delta_{\lambda,-}+a_{eR}\delta_{\lambda,+}\over
2s^2_W(s-m^2_Z)} \right ]\beta_W(1-2\lambda\mu'\cos\theta)~~. \label{FBorn-LT}
\eqa

The amplitudes in (\ref{FBorn-TL}, \ref{FBorn-LT}) are both HV, and at high energies
they are quickly  suppressed like $m_W/ \sqrt{s}$.\\

\noindent
The {\bf Longitudinal-Longitudinal (LL) amplitudes  ($\mu=0,\mu'=0$)}
are
\bqa
F^{\rm Born}_{\lambda 00}&=& {se^2\sin\theta\over16ts^2_W}\delta_{\lambda,-}
\left \{ {s\over m^2_W}(\beta_W-\cos\theta)+2\beta_W \right\}
\nonumber\\
&&+{(2\lambda)s^2e^2\over8m^2_W}
\left [{Q_e\over s}+{a_{eL}\delta_{\lambda,-}+a_{eR}\delta_{\lambda,+}\over
2s^2_W(s-m^2_Z)} \right ]\beta_W(3-\beta_W^2)\sin\theta~~, \label{FBorn-LL}
\eqa
where (\ref{kinematics}) have again been used.
At high energy, keeping terms to order  $m^2_Z/ s$ and $m^2_W/ s$, one gets
\bqa
F^{\rm Born}_{-{1\over2}00} & \to & - {e^2\over8s^2_Wc^2_W} \sin\theta ~~,
\nonumber \\
F^{\rm Born}_{+{1\over2}00} & \to & {e^2\over4c^2_W} \sin\theta ~~, \label{Born-asym-LL-HC}
\eqa
which together with (\ref{Born-asym-TT-HC}) confirm that all Born HC  amplitudes in
(\ref{4HC-amp-list}), go to constants, asymptotically.
On the contrary,  all six HV amplitudes listed in (\ref{HV-amp-list})  vanish,
 in this limit.\\

\begin{figure}[p]
%\vspace{-1cm}
\[
\epsfig{file=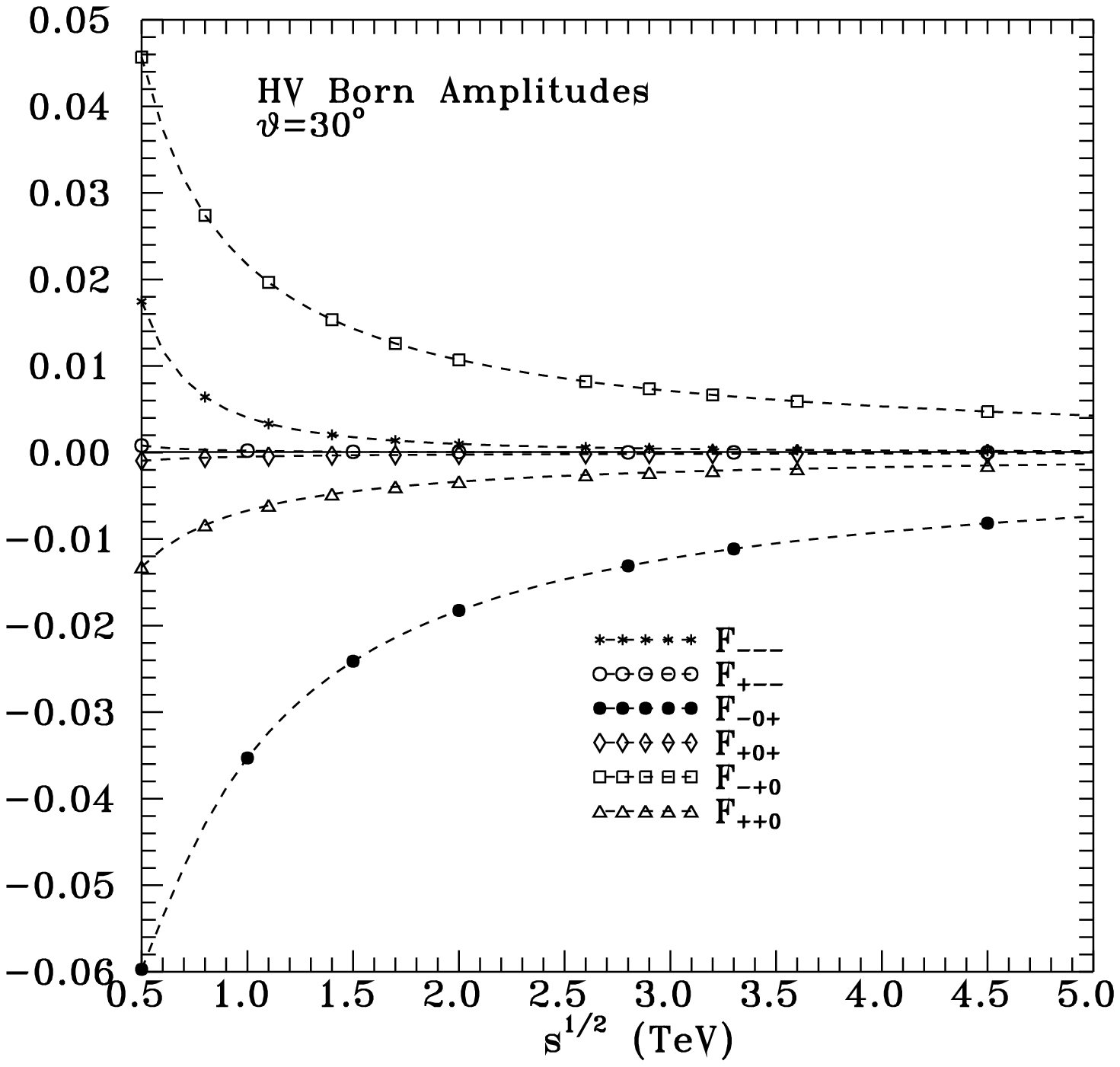, height=6.cm}\hspace{1.cm}
\epsfig{file=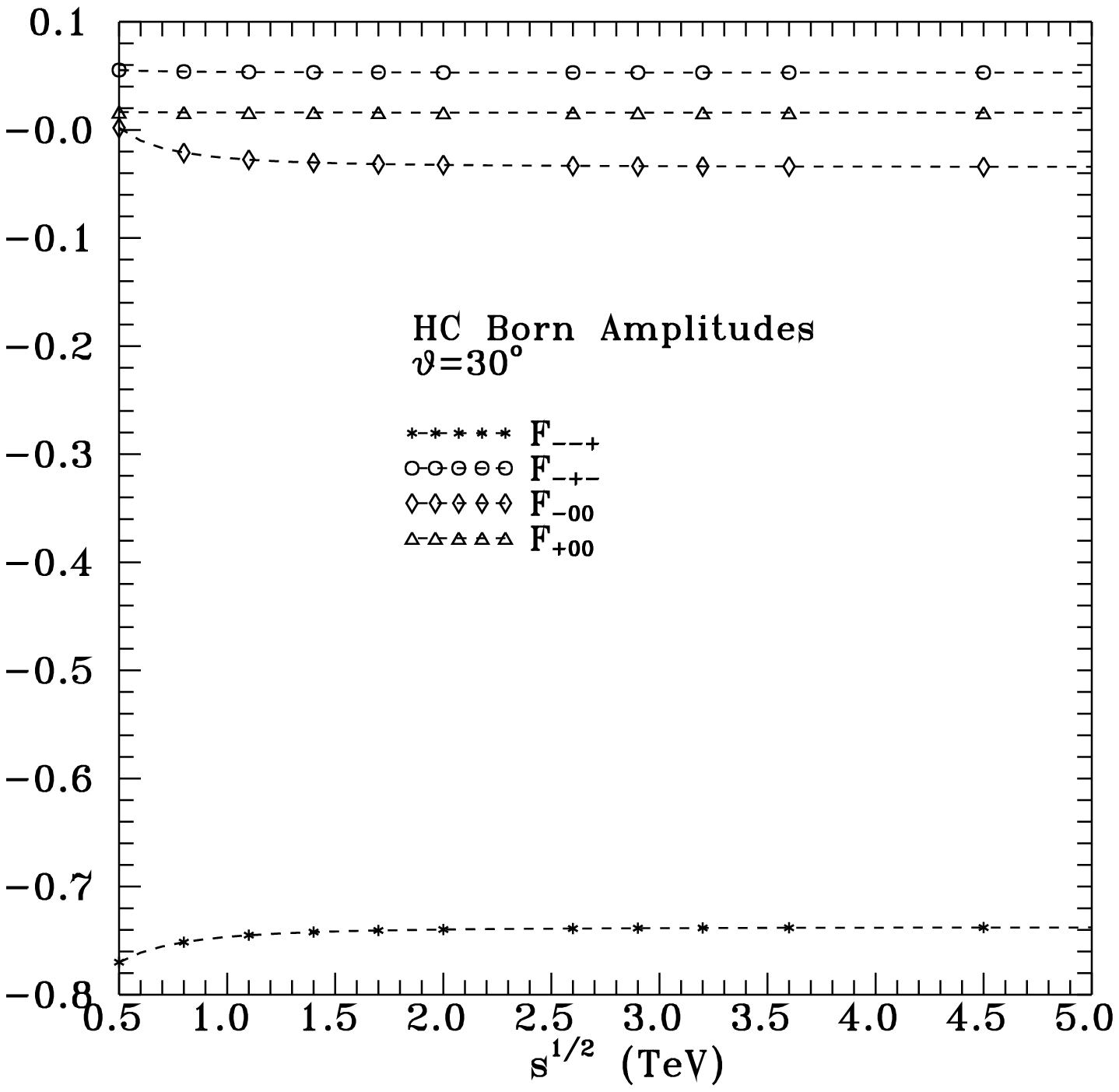,height=6.cm}
\]
\caption[1]{Left panel: Born contributions to the six helicity violating (HV)
amplitudes listed in (\ref{HV-amp-list}).
Right panel:  Born contributions to the  the four helicity conserving (HC)
 amplitudes listed in (\ref{4HC-amp-list}).}
\label{HV-HC-Born-amp}
\end{figure}

The Born level  properties of the helicity amplitudes are illustrated
in Figs.\ref{HV-HC-Born-amp}. The two  HC amplitudes listed in (\ref{R-HC-amp-list}),
are not shown, since they vanish, when coefficients proportional to  the
electron-mass are neglected.\\

\subsection{ Helicity amplitudes to the 1loop  electroweak (EW) order.}

The relevant contributions come from  up and down triangle diagrams in the t-channel;
initial and
final triangle diagrams in the s-channel; direct, crossed and
twisted box diagrams;  specific triangles involving a 4-leg gauge boson
couplings; and finally neutrino, photon and Z self-energies.
Counter terms in the Born contributions, which help canceling the
divergences induced by  self-energy and triangle diagrams,  are also included, leading to the so-called
on-shell renormalization scheme \cite{OS}.

Such type of computations have already been done; see for example \cite{Hahn, Denner}.
But our aim here is  to look at the specific properties of each helicity
amplitudes, and to derive simple high energy expressions for the
HC ones. For this reason we repeated the complete calculation of the 1loop EW corrections
and then  computed their high energy expressions that we call supersimple (sim),
using the expansion of \cite{asPV}. A special attention is paid to the
virtual photon exchange diagrams leading to infrared singularities (when $m_\gamma\to 0$)
which are then cancelled by the addition of the soft photon bremsstrahlung contribution. The sim
expressions are given (in Appendix A) in the two possible choicess , arbitrary small $m_\gamma$
value, or $m_\gamma=m_Z$ which can be considered as "small" at high energies. This second
choice, also used in previous studies \cite{super, ttbar}, has the advantage of
leading to even simpler expressions as we can see in Appendix A.\\

As already said and numerically shown below, the HV amplitudes amplitudes in (\ref{HV-amp-list})
 are negligible at high energies. Only the four HC amplitudes appearing in (\ref{4HC-amp-list})
 are relevant there. Turning to them, we present in Appendix A.1
 the very simple   sim expressions
 for the  TT amplitudes  $F_{--+}$, ~$F_{-+-}$; while the corresponding expressions for the LL
amplitudes $F_{-00}$,~ $F_{+00}$ appear  in Appendix A.2.
The results (\ref{simSM--+}, \ref{simSM-+-}, \ref{simSM+00}, \ref{simSM-00})
give the SM  predictions,
while  (\ref{simMSSM--+}, \ref{simMSSM-+-}, \ref{simMSSM+00}, \ref{simMSSM-00})
give the MSSM ones, always corresponding to the   $m_\gamma=m_Z$ choice.
The corrections to be done to them in order to  obtain the general result for any  $m_\gamma$,
appear in   (\ref{deltaFTT}, \ref{deltaF00}).

For  deriving these, we start  from the  complete  1loop EW results
in terms of Passarino-Veltman (PV) functions \cite{PV},
and then  use their high energy expansions given in  \cite{asPV}.
For  the  TT amplitudes  $F_{--+}$, ~$F_{-+-}$, the derivation is quite
straightforward.

For the two LL amplitudes $F_{-00}$, $F_{+00}$ though, the derivation  is very delicate,
because of huge   gauge cancelations
among contributions  exploding  like\footnote{ Particularly for neutralinos, this demands
a very accurate determination of their mixing matrices, like the one supplied  e.g. by  \cite{LeMouel}.}
  $s/ m^2_W$. Such cancelations also occur  at Born level,
between t- and s-channel terms. But at 1loop level, the situation  is
much more spectacular, because  more  diagrams are involved.
Technically, the derivation of the limiting  expressions can be facilitated  by using
the equivalence theorem and  looking at the Goldstone process $e^-e^+\to G^-G^+$
\cite{equivalence}.\\

We next turn to the infrared divergencies implied by the presence of
 $m_\gamma$ in the $e^-e^+\to W^-W^+$ amplitudes. As usual, these are  canceled at the cross section
level by adding to the 1loop EW results for  $d\sigma (e^-e^+\to W^-W^+)/d\Omega$, the Born-level
 cross section describing the soft photon bremsstrahlung, given by
\bq
\frac{d\sigma_{\rm brems}(e^-e^+\to W^-W^+\gamma)}{d\Omega}=
 \frac{d\sigma^{\rm Born} (e^-e^+\to W^-W^+)}{d\Omega} \delta_{\rm brems}(m_\gamma,\Delta E)~~,
 \label{brems-sigma}
\eq
where $\delta_{\rm brems}(m_\gamma,~\Delta E)$ is given by\footnote{Parameter $\lambda$ in \cite{Boehm} corresponds  to our $m_\gamma$} Eqs. (5.18) in \cite{Boehm}, while
 $\Delta E$   describes the highest  energy  of the emitted unobservable
soft photon, satisfying
\bq
m_\gamma  \leq \Delta E \ll \sqrt{s} ~~. \label{brems-kin}
\eq
The only  requirement for this cancelation  to happen is that
  $m_\gamma$ is {\it small}; i.e. that terms proportional to a power of  $m_\gamma$
  (not inside a high energy logarithm) are always negligible. Under these condition, any
  $m_\gamma$-dependence cancels out in the sum $d\sigma (e^-e^+\to W^-W^+)/d\Omega$ plus
  $d\sigma_{\rm brems}/d\Omega$.

 But, at the high energies of $\sqrt{s} \gg m_Z$ we are interested in,
 the $Z$ mass is also {\it small}; since
 any such $m_Z$ coefficient is necessarily suppressed by an energy denominator.  In other words,
 since the infrared $m_\gamma$ effects cancel out in the cross section including  bremsstrahlung (\ref{brems-sigma}) contribution, they will also cancel in the special case  $m_\gamma=m_Z$.
 As already said we made this choice because it leads to the simplest expressions.
 The illustrations given below correspond to it.

 In  order to obtain the (infrared sensitive) unpolarized cross section
 $d\sigma (e^-e^+\to W^-W^+)/d\Omega$ from the experimental data, one has obviously to  subtract the
  bremsstrahlung contribution. Consequently, the difference between  the values of this cross section
  regularized  at an arbitrary $m_\gamma$ or  at  $m_\gamma=m_Z$, for the same $\Delta E$,
   is  given by
  \bqa
 && \frac{d\sigma (e^-e^+\to W^-W^+) }{d\Omega}\Big |_{m_\gamma} -
 \frac{d\sigma (e^-e^+\to W^-W^+) }{d\Omega}\Big |_{m_\gamma \to m_Z} \nonumber \\
   && = \frac{d\sigma^{\rm Born}}{d\Omega}
     {\alpha\over\pi}\left ( \ln {m_Z\over m_\gamma} \right )
  \left ( 4 -2\ln{s\over m^2_e} +4\ln{m^2_W-u \over m^2_W-t}
 +2{s-2m^2_W\over s\beta_W}\ln{1-\beta_W\over 1+\beta_W} \right ) ; \label{mgamma-mZ-effect}
 \eqa
 see our eqs.(\ref{kinematics}, \ref{brems-sigma}) and  eq.(5.18) of \cite{Boehm}.
 If one wants to keep the usual choice of an arbitrary  small $m_\gamma$ in the bremsstrahlung cross section,
 one would have to use our extended sim expressions given in (\ref{deltaFTT}, \ref{deltaF00})
 of Appendix A.\\

Turning now to the numerical illustrations, we first check
that  all HV amplitudes quickly vanish at high energy, in both MSSM and SM  \cite{heli1, heli2}.
For the  MSSM case, we use benchmark  S1 of \cite{bench}, where the EW
 scale values of  all squark masses are at the 2 TeV level, $A_t=2.3$ TeV,
 the slepton  masses are at $0.5$ TeV, and the remaining mass parameters (in TeV)  are
 \bq
 \mu =0.4~~,~~ M_1=0.25 ~~,~~ M_2=0.5 ~~,~~ M_3=2 ~~, \label{bench-param}
 \eq
while $\tan\beta=20$.
Such a benchmark  is  consistent with present LHC constraints \cite{bench}.
All MSSM results shown in this paper, are using this benchmark.
Similar results are also obtained for other LHC-consistent  MSSM benchmarks, like those
listed e.g. in  the Snowmass suggestion \cite{Snowmass},
or the very encouraging  cMSSM ones given  in \cite{Konishi}.

\begin{figure}[p]
%\vspace{-1cm}
\[
\epsfig{file=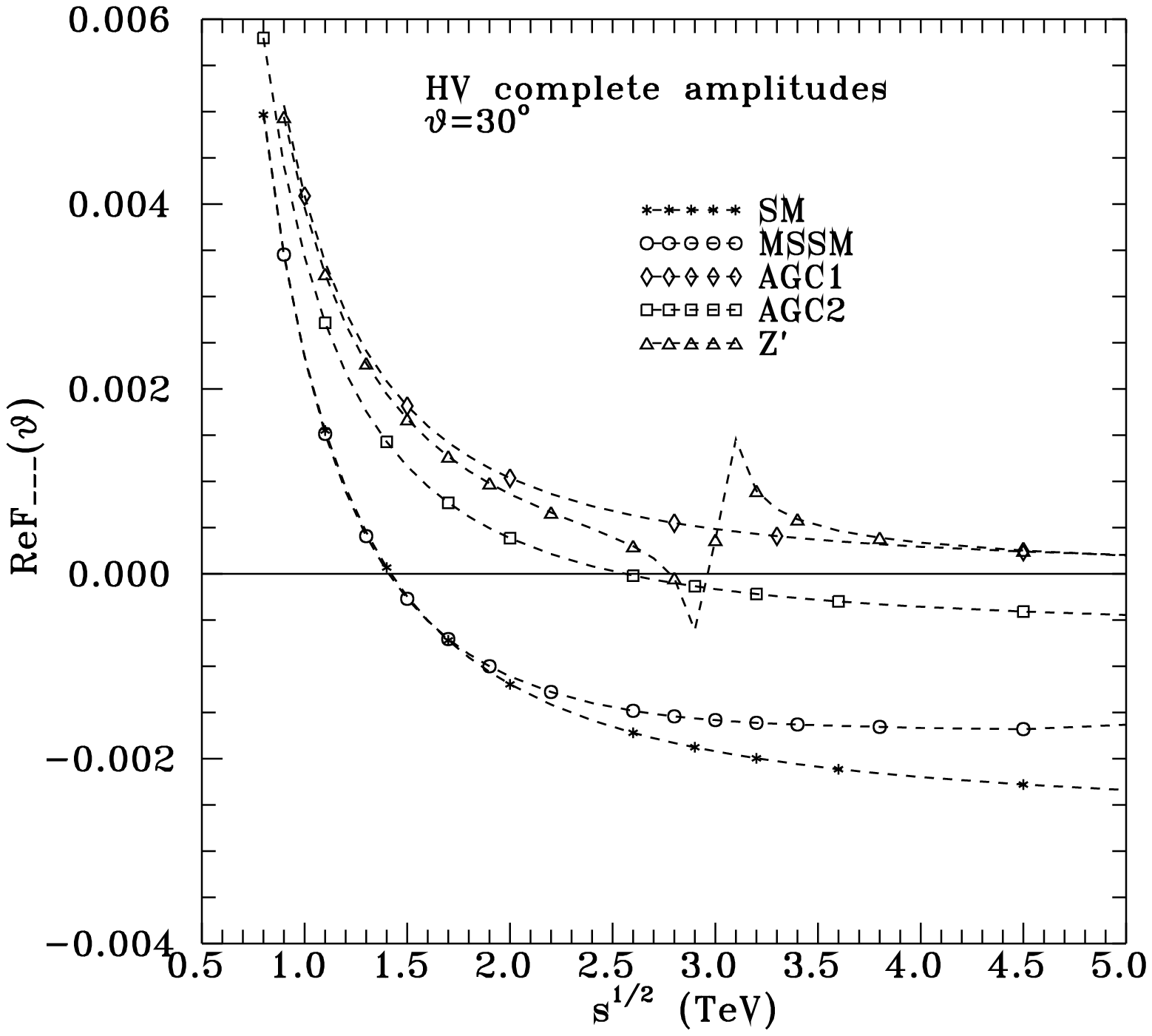, height=6.cm}\hspace{1.cm}
\epsfig{file=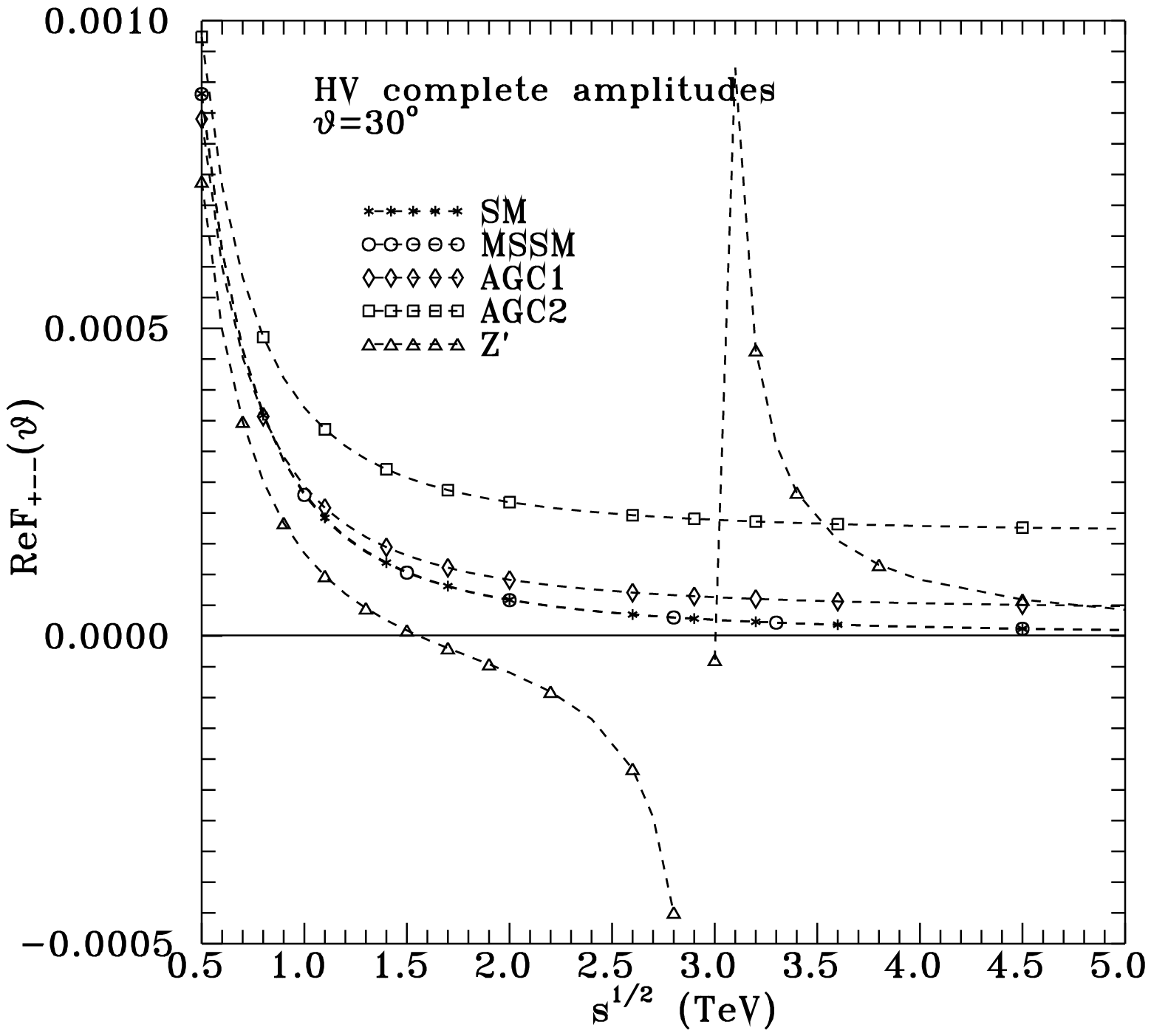,height=6.cm}
\]
\[
\epsfig{file=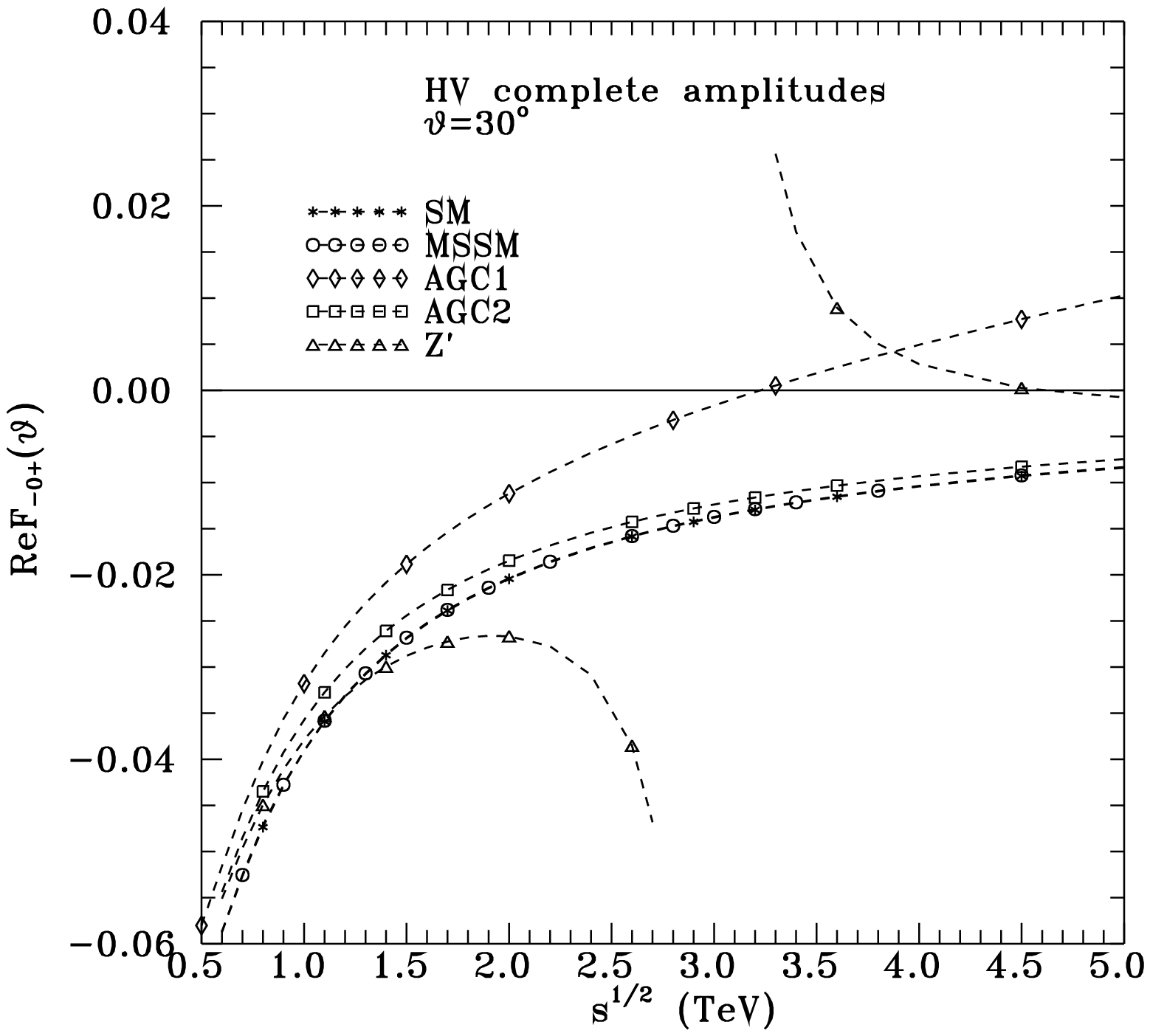, height=6.cm}\hspace{1.cm}
\epsfig{file=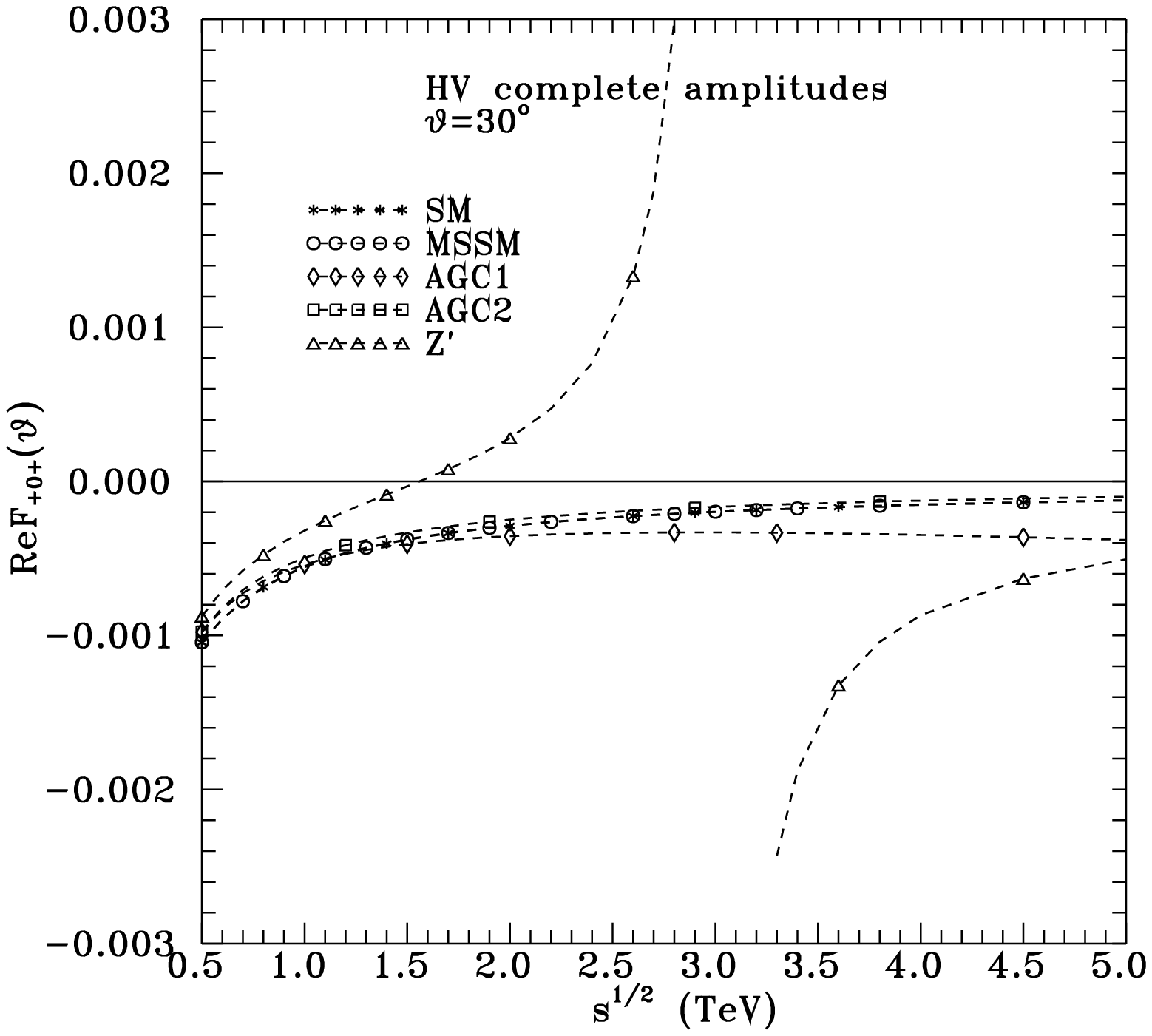,height=6.cm}
\]
\[
\epsfig{file=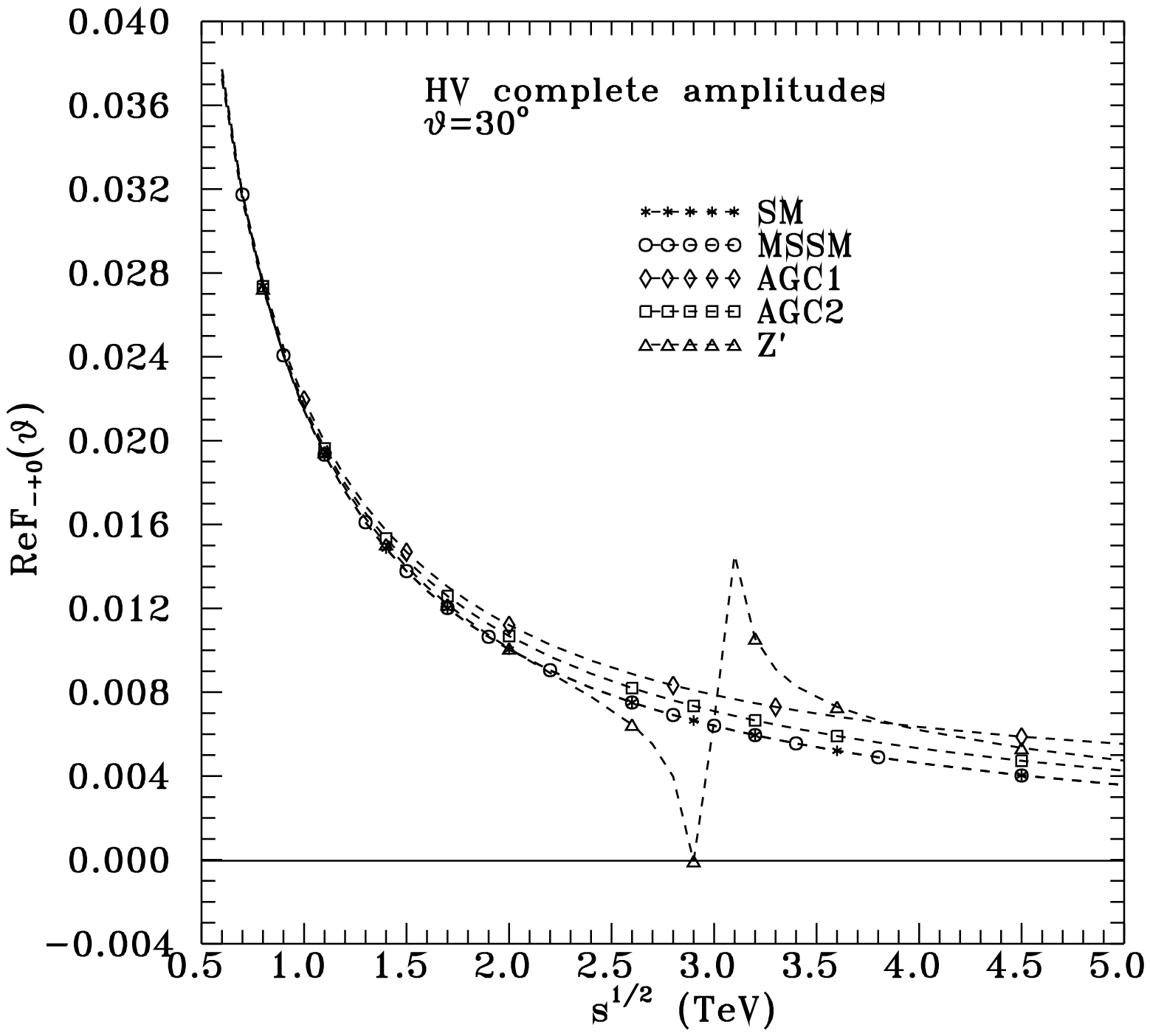, height=6.cm}\hspace{1.cm}
\epsfig{file=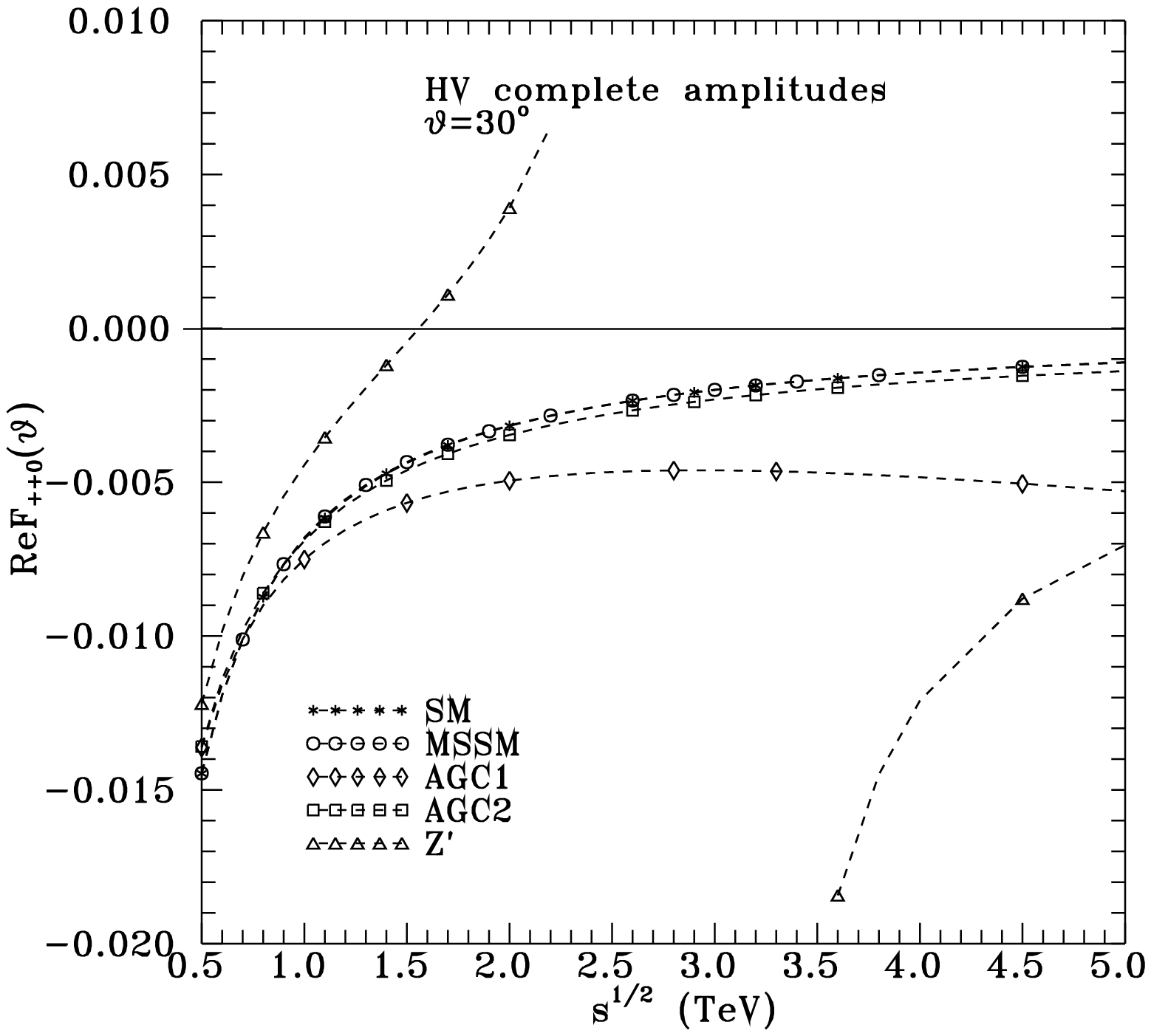,height=6.cm}
\]
\caption[1]{The real parts of the six HV amplitudes listed in  (\ref{HV-amp-list}),
at 1loop EW order in SM and MSSM, using the $m_\gamma=m_Z$ regularization.
The New Physics contributions  from
AGC1, AGC2, or  a new $Z'$ (see text) are also shown. The horizontal solid lines
indicate  how these HV amplitudes compare to  a vanishing
 asymptotic value expected in MSSM. The imaginary parts of the amplitudes are much smaller,
 since they receive no Born contribution.}
\label{HV-full-amp}
\end{figure}

Comparing the SM and MSSM results in Figs.\ref{HV-full-amp},
we see that  for all HV amplitudes,  the purely supersymmetric
contribution mostly cancel the (already suppressed)
pure SM ones; this is more spectacular for energies above the SUSY
scale. Thus, Figs.\ref{HV-full-amp} indeed show that the  six HV amplitudes listed
in  (\ref{HV-amp-list}),  are quickly suppressed in MSSM, as well as  in SM.\\

\begin{figure}[p]
%\vspace{-1cm}
\[
\epsfig{file=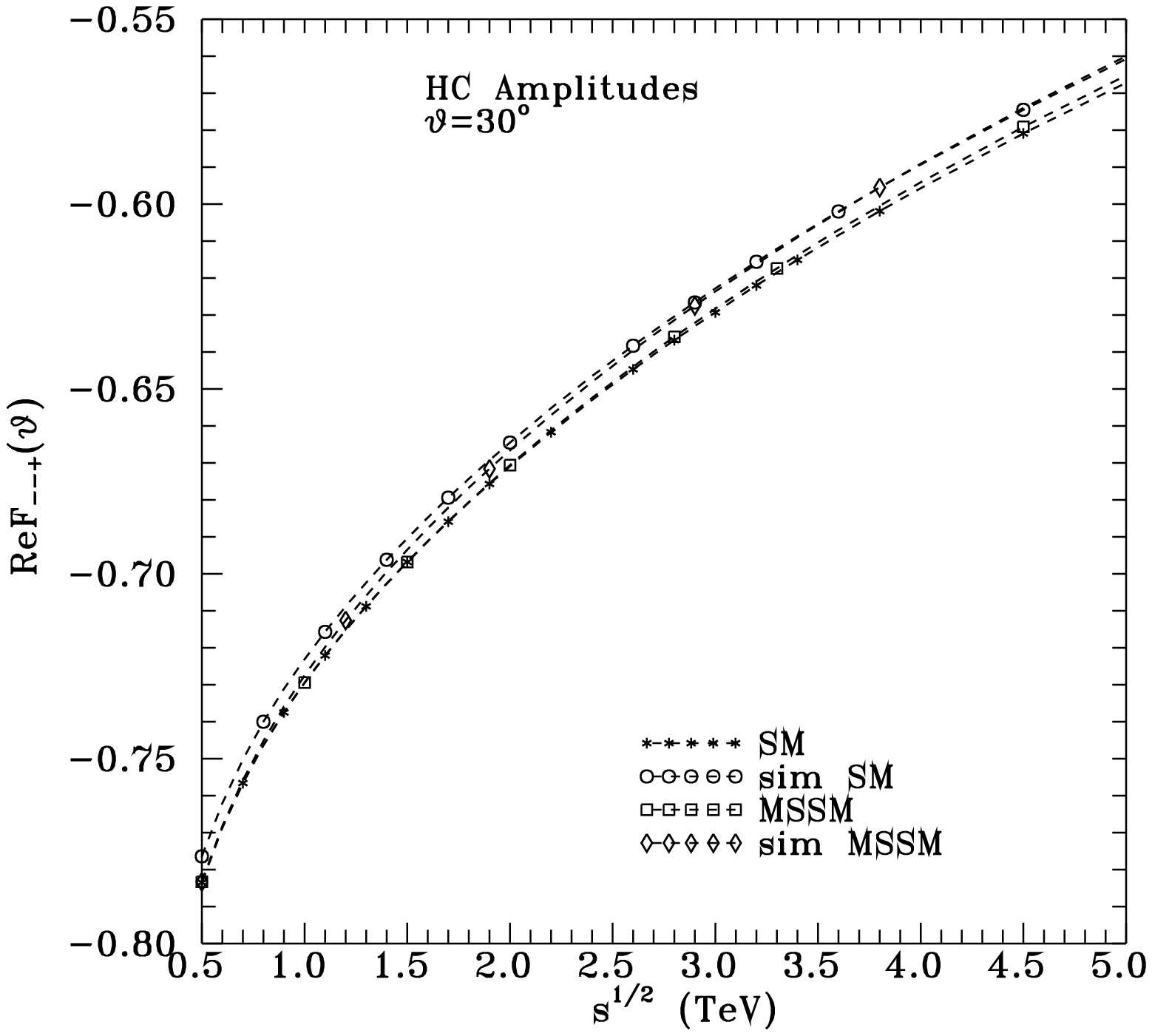, height=6.cm}\hspace{1.cm}
\epsfig{file=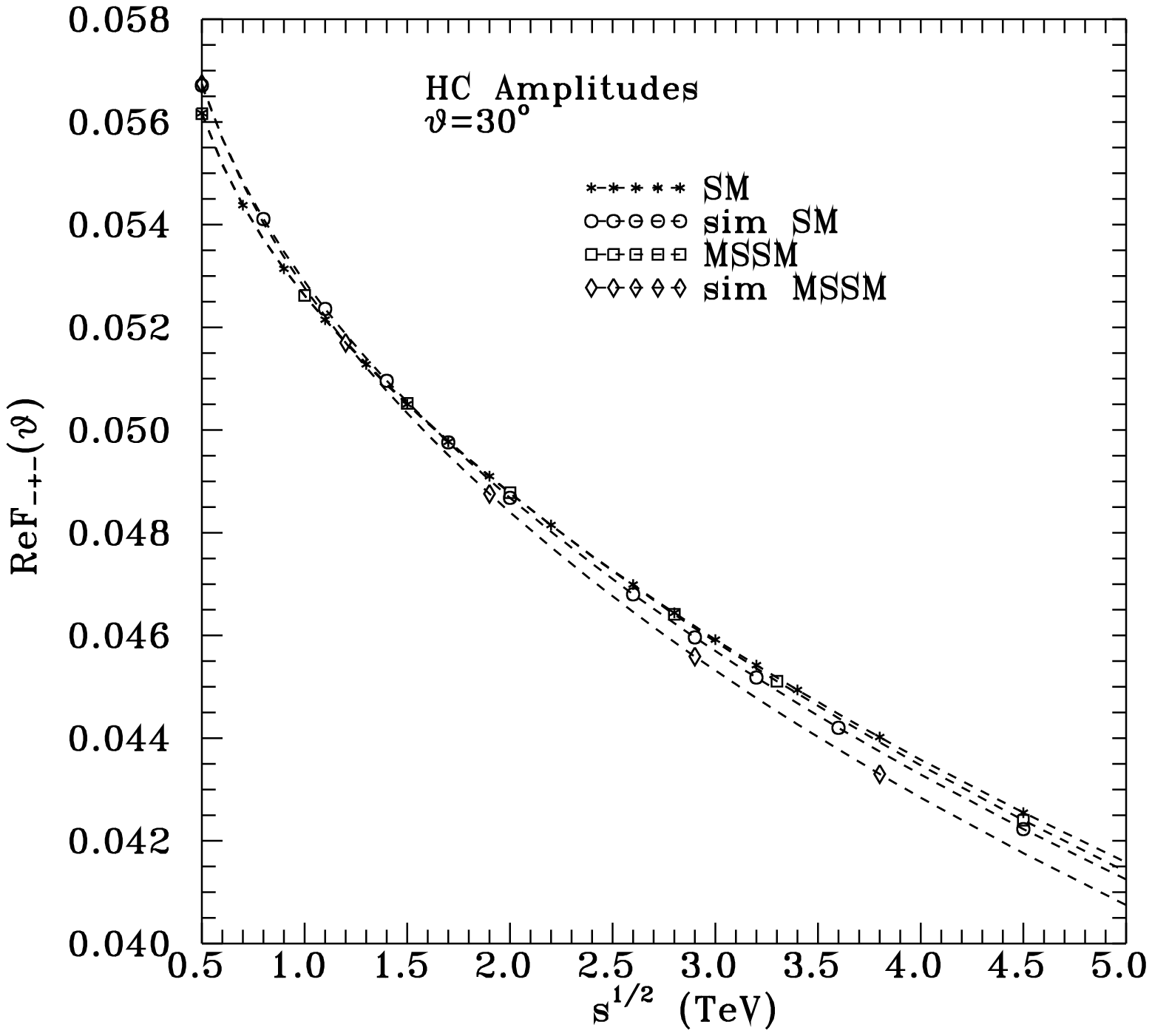,height=6.cm}
\]
\[
\epsfig{file=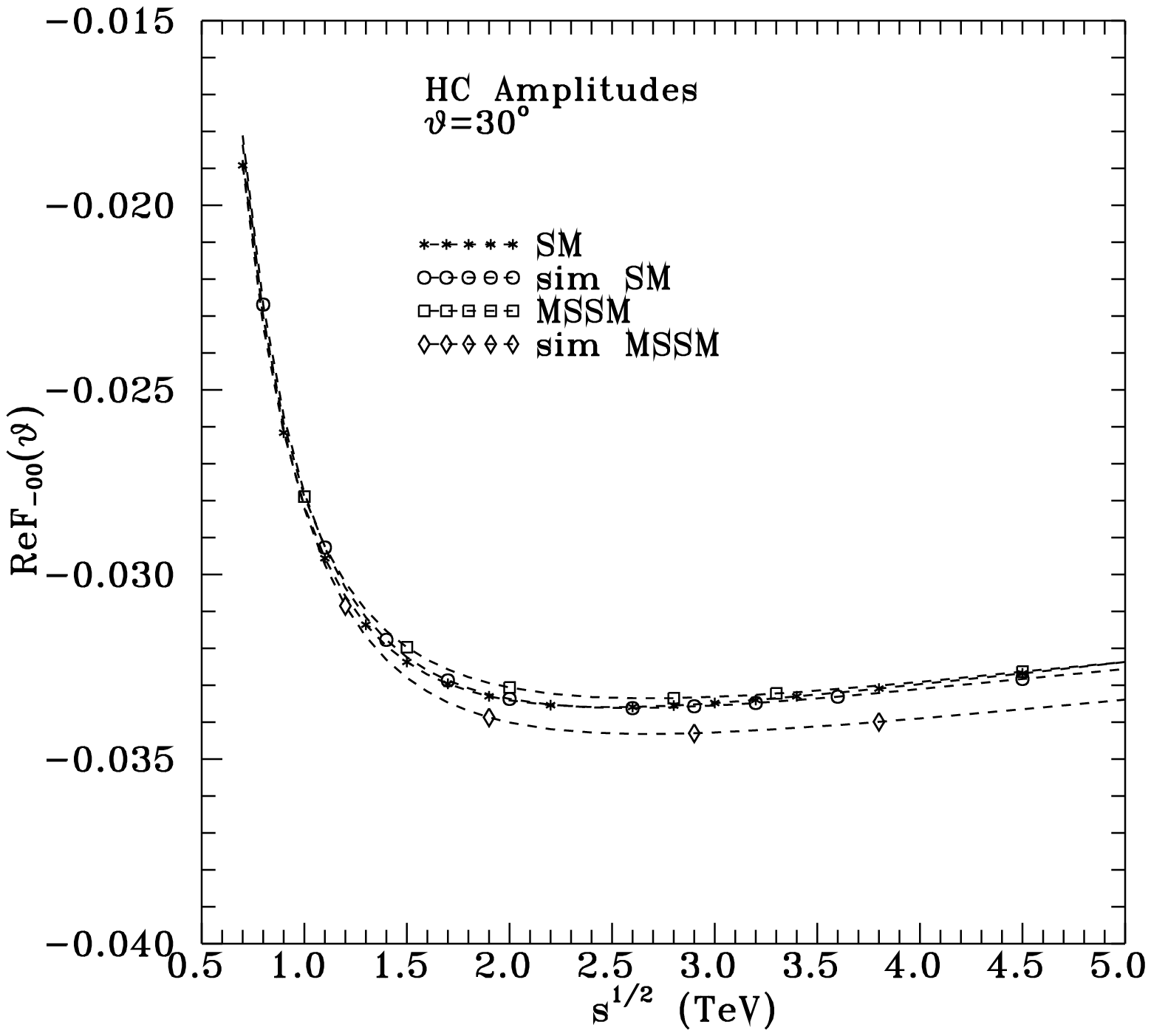, height=6.cm}\hspace{1.cm}
\epsfig{file=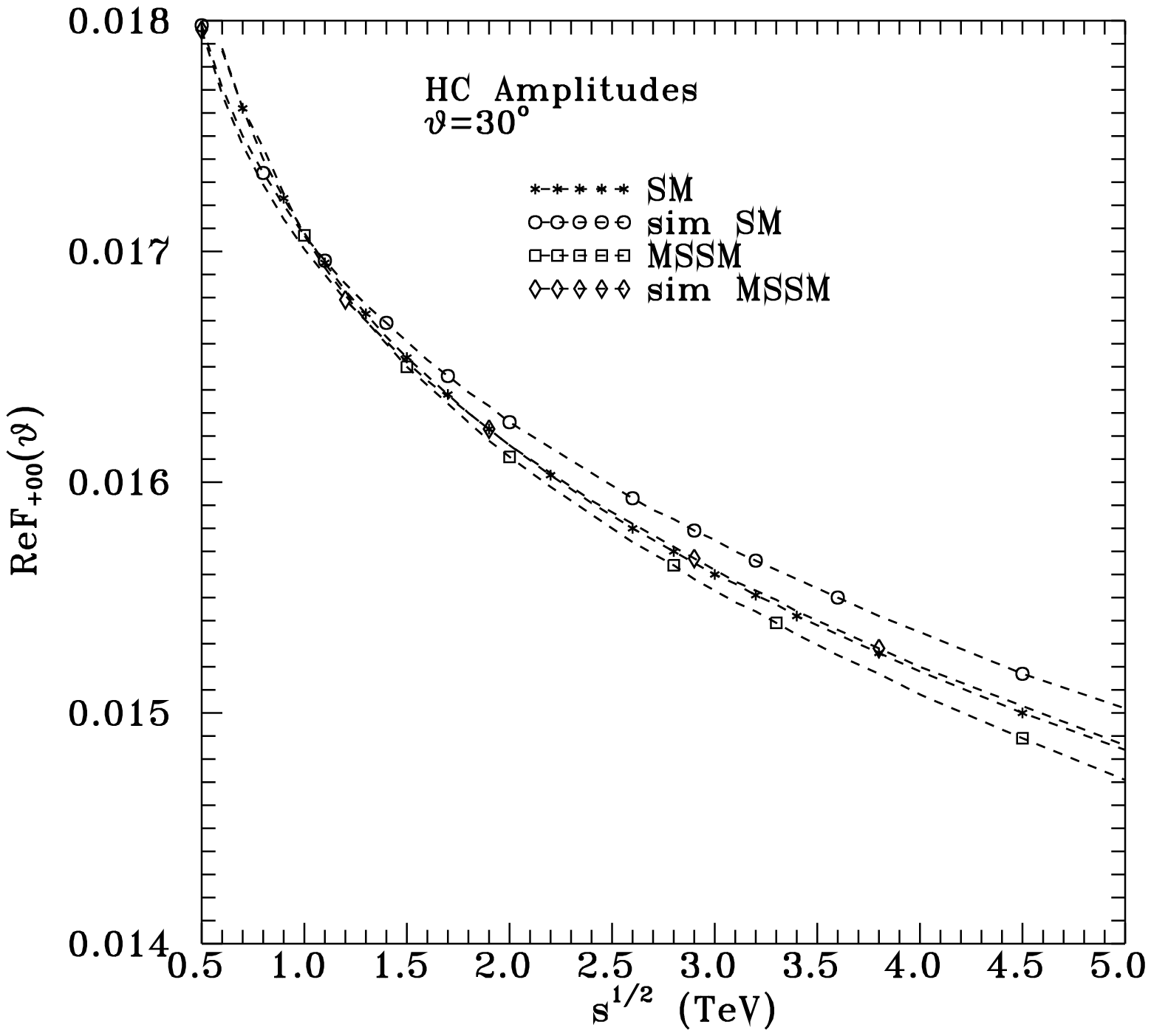,height=6.cm}
\]
\caption[1]{The real parts of the  complete  1loop EW results for
the four HC amplitudes listed in
(\ref{4HC-amp-list}), and their  supersimple (sim) approximations,
in  SM and the MSSM benchmark described in the text.
Upper (lower) panels describe the TT (LL)
amplitudes respectively. The imaginary parts of the amplitudes are much smaller,
 since they receive no Born contribution.}
\label{HC-full-amp}
\end{figure}

We next turn to the high energy description of the four
leading (HC) amplitudes listed in (\ref{4HC-amp-list}).
 As it is shown in Figs.\ref{HC-full-amp}, the  supersimple  (sim)  approximations to them,
 follow very closely the complete expressions for the  1loop electroweakly  corrected
 amplitudes, in both SM and MSSM.
For the TT amplitudes $F_{--+}$, $F_{-+-}$, this appears in the upper panels
of Figs.\ref{HC-full-amp}, for  SM and the MSSM benchmark mentioned above.
The corresponding numerical illustrations for the  LL HC amplitudes are shown in the lower
panels. These results indicate that all four  1loop predictions; i.e. the complete SM and
MSSM results, as well their  sim SM and sim MSSM approximations,
are very close to each other at high energies. Moreover, a comparison of
Figs.\ref{HV-full-amp} and \ref{HC-full-amp} immediately shows that soon above 0.5TeV
the HC amplitudes in (\ref{4HC-amp-list}) are much larger than all other ones.

There are two main conclusions we  draw from this, for energies up to a TeV or so:
The first is that the process
 $e^-e^+\to W^-W^+$ is rather insensitive to MSSM contributions, for benchmarks consistent
with the present SUSY constraints \cite{bench, Snowmass, Konishi}. And the second conclusion
is that (\ref{simSM--+}, \ref{simSM-+-}, \ref{simSM+00}, \ref{simSM-00})
provide a true description of the sources of the relevant dynamics.\\

\section{Application to the $e^-e^+\to W^-W^+$ observables}

 The observables we  study here are  the unpolarized differential cross sections
\bq
{d\sigma\over d\cos\theta}={\beta_W\over 128\pi s}
\Sigma_{\lambda \mu \mu'}|F_{\lambda \mu \mu'}(\theta)|^2 ~~, \label{dsigma-unpol}
\eq
as well as the polarized differential cross sections using  right-handedly polarized
 electron beams $e^-_R$,
\bq
{d\sigma^R\over d\cos\theta}={\beta_W\over 64\pi s}
\Sigma_{\mu \mu'}|F_{+{1\over2},\mu \mu'}(\theta)|^2 ~~, \label{dsigma-pol}
\eq
where (\ref{kinematics}) is used.

\begin{figure}[p]
%\vspace{-1cm}
\[
\epsfig{file=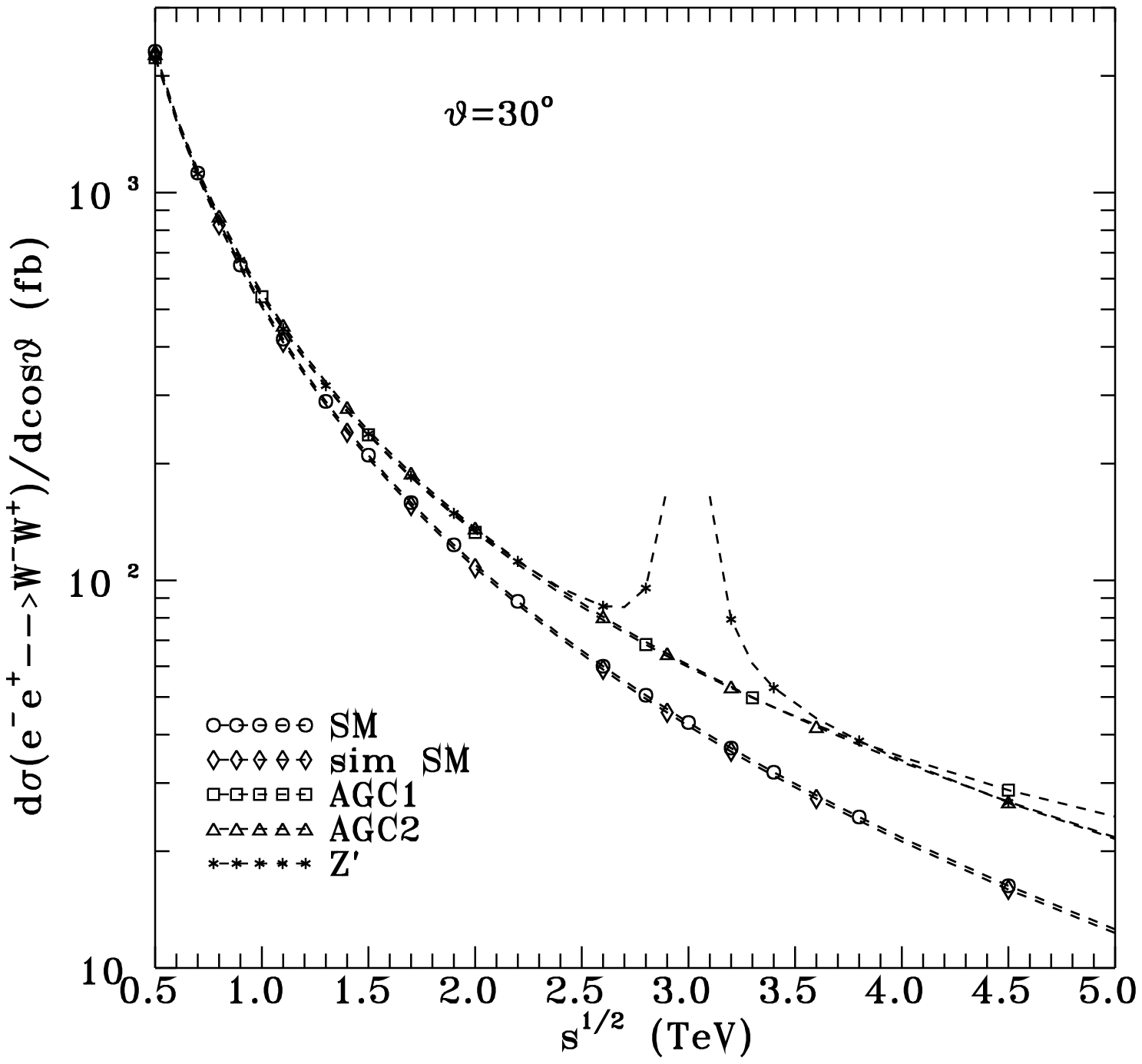, height=6.cm}\hspace{1.cm}
\epsfig{file=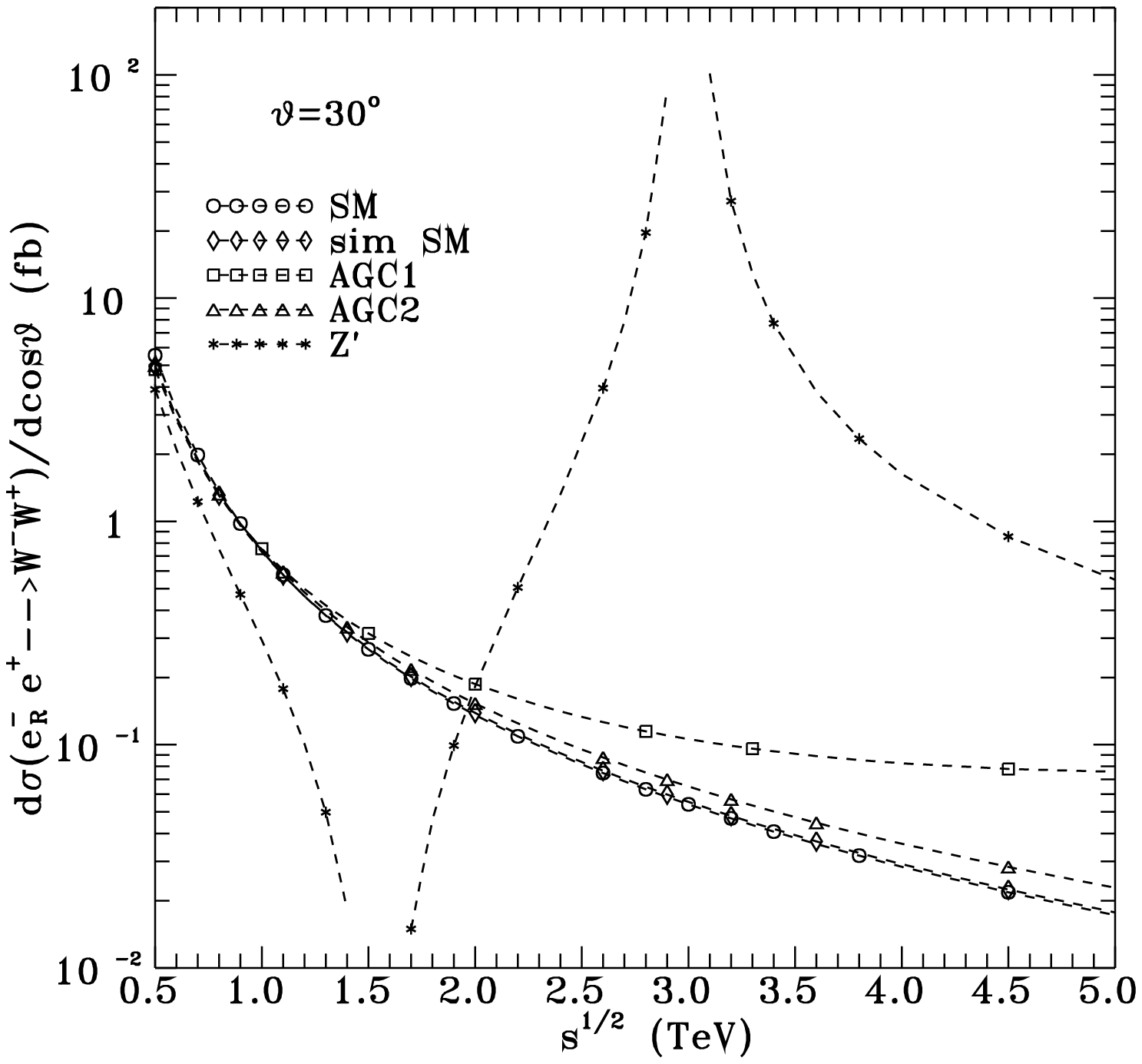,height=6.cm}
\]
\[
\epsfig{file=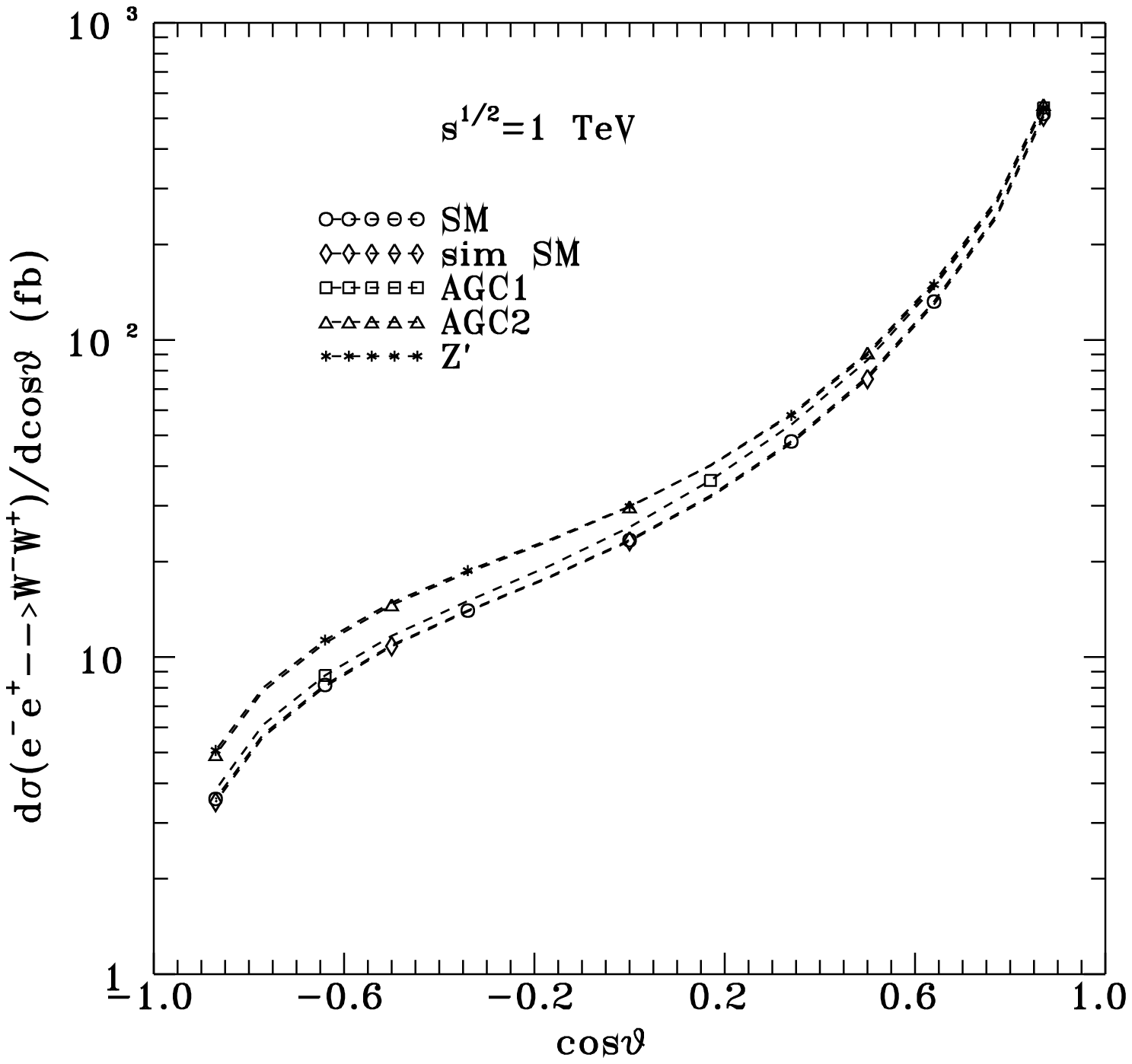, height=6.cm}\hspace{1.cm}
\epsfig{file=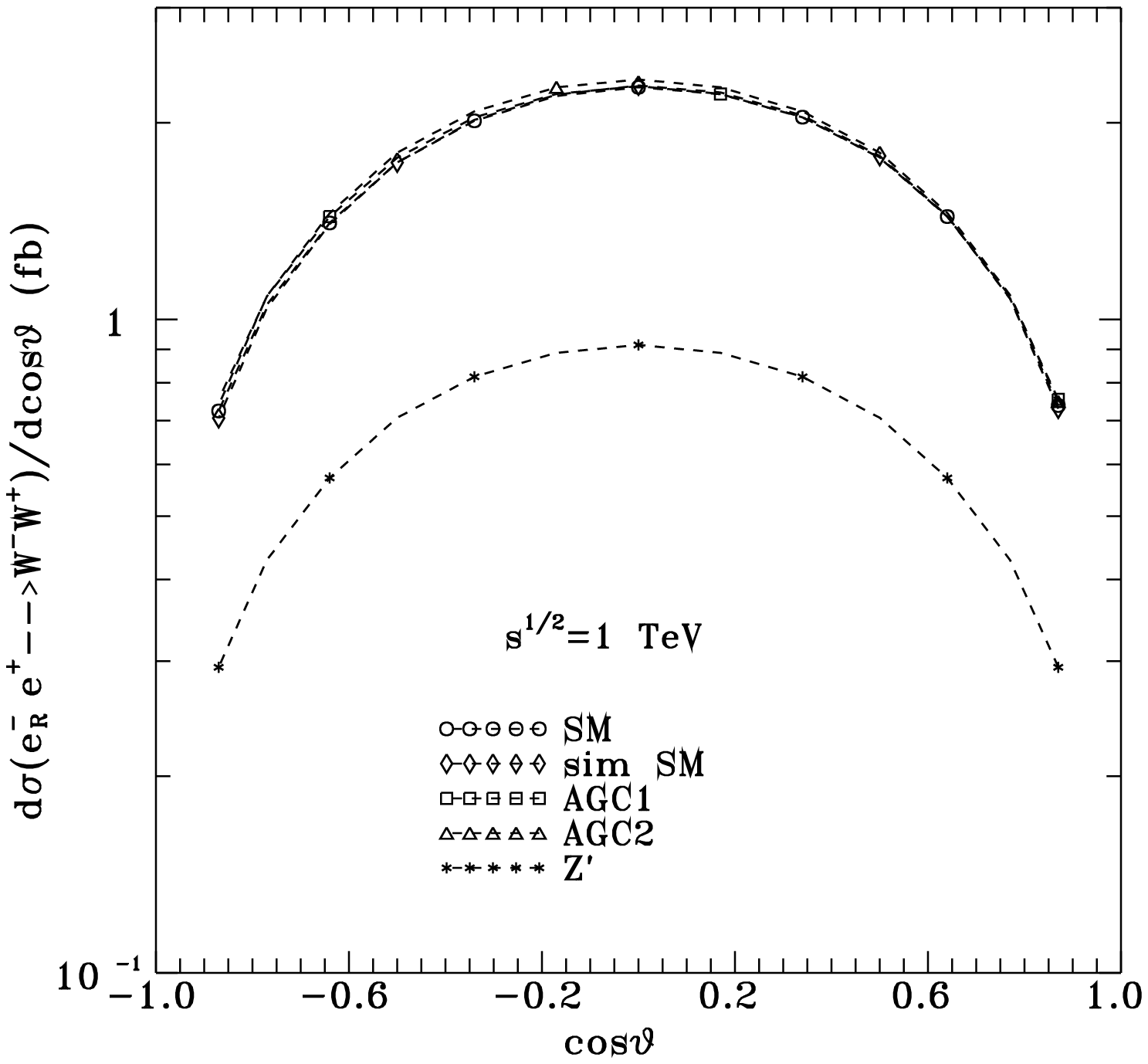,height=6.cm}
\]
\[
\epsfig{file=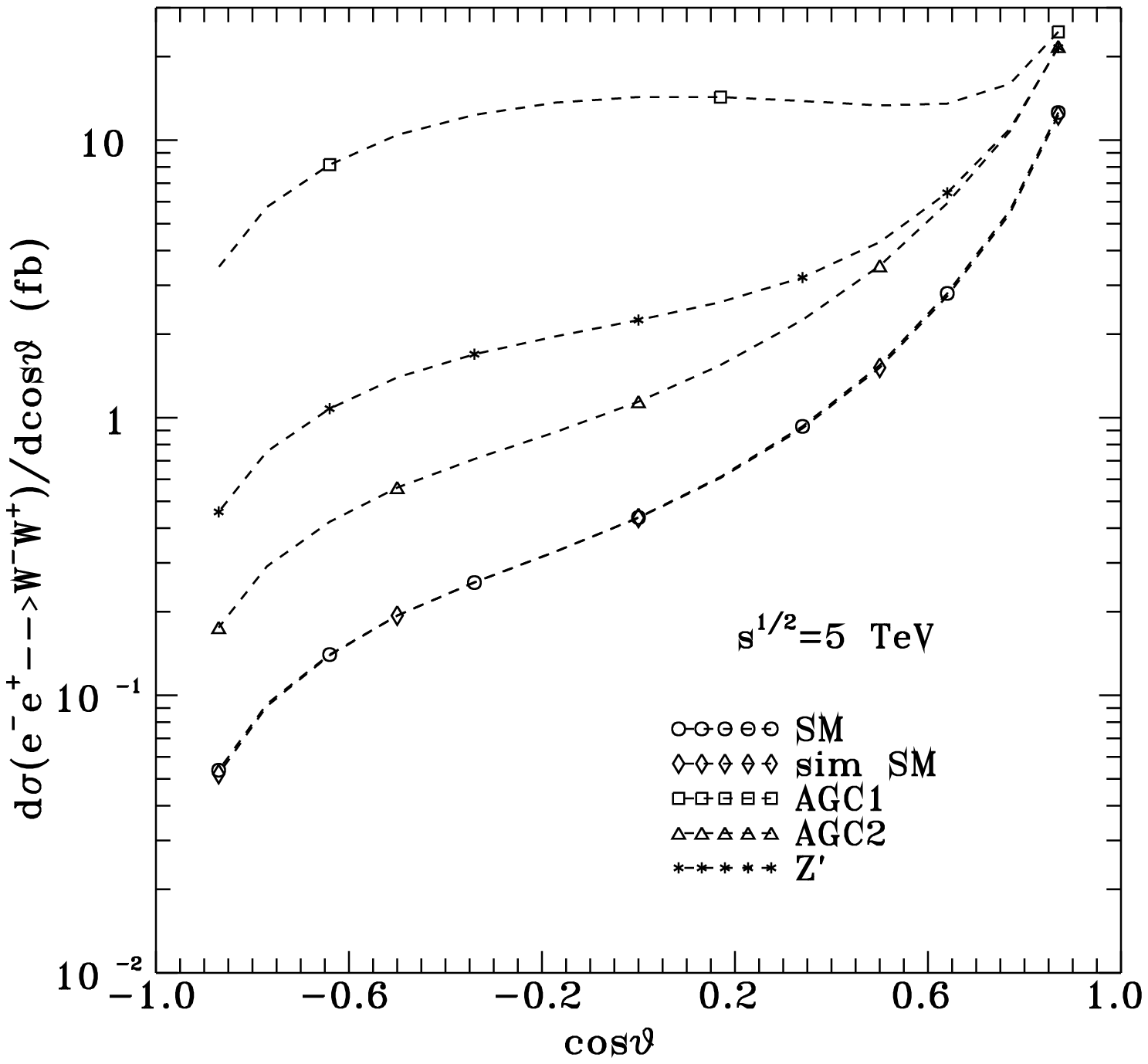, height=6.cm}\hspace{1.cm}
\epsfig{file=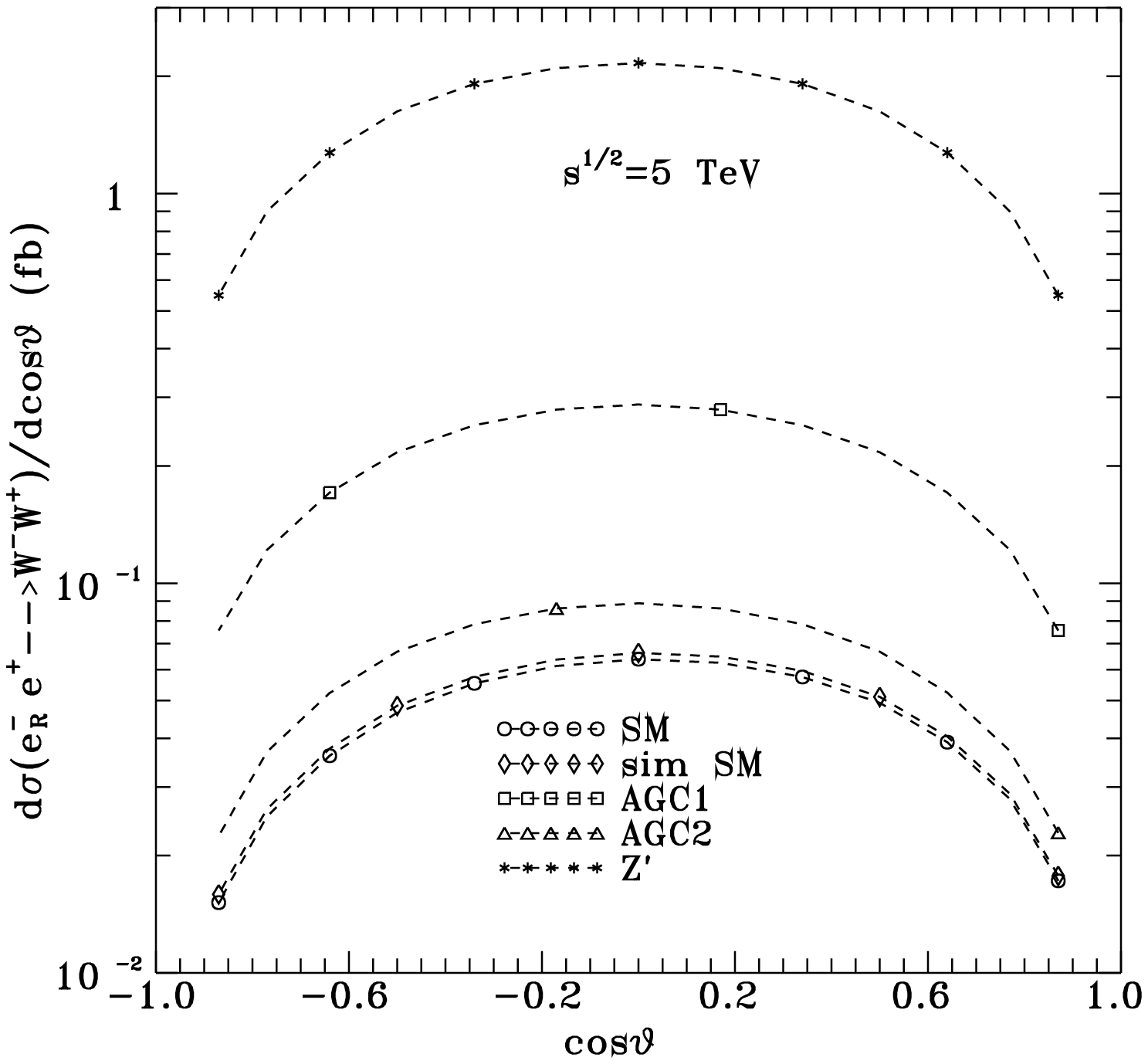,height=6.cm}
\]
\caption[1]{Differential cross sections for unpolarized $e^-e^+$ (left panels),
and right-electron polarized $e^-_Re^+$ (right panels),
in SM, sim SM and some New Physics models (see text).
Upper panels show the energy dependencies at $\theta=30^o$. Middle (lower) panels give
the angular
dependencies at $\sqrt{s}=1$ TeV ($\sqrt{s}=5$ TeV).}
\label{sigmas}
\end{figure}

These cross sections are shown in Figs.\ref{sigmas}, where the complete
1loop EW order SM results are  compared to  the corresponding
supersimple (sim) ones. The later are   constructed
by using the expressions of Appendix A for the HC amplitudes, while  the HV amplitudes
are approximated by  the quickly vanishing Born contributions\footnote{If instead
we had completely ignored the HV amplitudes in the sim cross sections, then
appreciable differences would only appear  for energies  below 1TeV,
particularly  for the $e^-_R$ cross sections.} in
(\ref{FBorn-TT}, \ref{FBorn-TL}, \ref{FBorn-LT}). As shown in
Figs.\ref{sigmas}, the sim results very closely follow the SM  ones.

 In addition,   we show  in the same figures, how the complete 1loop   SM results are changed,
  when an anomalous  contribution is added like e.g.    AGC1 or AGC2, respectively
defined by (\ref{AGC1-choice}) or (\ref{AGC2-gLgR}, \ref{AGC2-choice}) of Appendix B.1; or
a $Z'$-effect defined Appendix B.2.

Left  panels in   Figs.\ref{sigmas} present results for
the unpolarized $e^-e^+$ cross sections; while right panels show results for the
$e^-_Re^+$ cross sections involving a right-handedly polarized  electron.

The upper panels present the energy dependencies at $\theta =30^o$;
while the middle (lower) panels indicate the angular dependencies at
$\sqrt{s}=1$TeV ($\sqrt{s}=5$TeV).

In all cases, the supersimple (sim) description is very good.
No MSSM or sim MSSM illustrations are given, since they are very close to the corresponding
 SM ones; at the 1-2\% level, for benchmarks consistent with the current LHC
 constraints \cite{bench, Snowmass, Konishi}.

 In other words, at the scale of  Figs.\ref{sigmas},
 the SM and MSSM results for \cite{bench}, would coincide.
Such a  weakness of the pure supersymmetric
 contributions,  has  been already noticed in previous analyses,
 \cite{Hahn}.  Because of the different
mass scales of the supersymmetric partners, at a given energy,
the absolute magnitudes of the SUSY 1loop effects
may differ notably. But relative to the SM contributions (Born + 1 loop),
they always remain very small.

Concerning the relevant dynamics for the unpolarized $e^-e^+$ cross sections,
we note that, at forward angles, they  are  dominated by the left-handed-$e^-$  TT
amplitudes.\\

For specific experimental studies of the LL amplitudes,
one can either make a final polarization analysis of the $W^\mp$-decays;
or use a right-handedly polarized  $e^-$-beam, so that
the usual  TT amplitudes  do not contribute.
In the right panels in Figs.\ref{sigmas}, we show the energy and angular dependencies
of these $e^-_R e^+$  cross sections.

 These LL studies are probably the best place to  search for
anomalous contributions, like those from the   AGC effects presented in Section B.1.
As seen in (\ref{FAGC-TT}-\ref{FAGC-LL}), such
AGC contributions   do not appear in the  HC TT amplitudes;
but they do appear in  the HC LL amplitudes, as well as
in all the  HV ones (TT, TL and LT).
This is a remarkable property that should be checked by a careful
analysis of experimental signals.

 The most simple-minded implication of    AGC physics is presented
 by the AGC1 model in Figs.\ref{HV-full-amp}, \ref{sigmas}, \ref{HC-full-NPamp},
 where  the   parameters in Appendix B.1 are fixed as  in  (\ref{AGC1-choice}).
 In this case, the  anomalous contributions to the LL
amplitudes  increase  like $ s/ m^2_W$,
causing   a strong  increase of the cross sections
with the energy.

Such a strong increase may be tamed though, by the existence of
 scales $M$ in the  various   anomalous couplings,  which transforms them to
  form factors  decreasing   like  $M^2/(s+ M^2)$.

Another way of taming the above strong AGC increase, is by the addition
of new exchanges in the
t-channel, such that one gets cancelations between s- and t-channel contributions,
 like in the Born SM case.
A  purely ad-hoc phenomenological solution of this kind is given by AGC2,  presented
in Appendix B.1, and determined by
(\ref{AGC2-gLgR}, \ref{AGC2-choice}). In the effective lagrangian framework
many such possibilities exist; see e.g. \cite{ef-Lagrangian}.

 The AGC1, AGC2 results of in Figs.\ref{HV-full-amp}, \ref{HC-full-NPamp}, \ref{sigmas},
show  various amplitudes  and cross-sections where such anomalous
behaviours may be seen and  compared to the SM and MSSM results.

Present experimental constraints on fixed AGC couplings,
 from LEP2  \cite{LEP2WW} are of the order of
 $\pm 0.04$.
 From LHC \cite{LHCWW}, they are of the order of $\pm 0.1$;
 compare with (\ref{AGC1-choice}, \ref{AGC2-choice}).
 These values are much larger than the uncertainties of
 our description.\\

Another type of anomalous contribution is a $Z'$ exchange in the s-channel;
 see \cite{Andreev} and Appendix B2.
 Here also one can impose a good high energy behaviour to the
LL and LT amplitudes. A simple solution is a $Z-Z'$ mixing
such that, the total s-channel contribution at high energy, cancels the standard
t-channel exchange at Born-level.
  Figs.\ref{HV-full-amp}, \ref{sigmas}, \ref{HC-full-NPamp}  show the
   behaviours of the various amplitudes
 and cross-sections under the presence of such $Z'$ contributions,
 and  compare them to the corresponding SM and MSSM ones.\\

>From the above illustrations one sees that our supersimple expressions
are sufficiently accurate  to distinguish 1loop SM or MSSM corrections
from such New Physics.   But these are examples.  More elaborated
analyses could of course  be done,
 for example in the  spirit of \cite{Andreev}; still remaining in a
 sensitivity region where  supersimple expressions sufficiently describe
SM physics. The existence of this possibility constitutes an important
 motivation for supersimplicity. \\

\section{Conclusions}

In this paper we have  analyzed the high energy behaviour
of the  1loop EW corrections to the
$e^-e^+\to  W^- W^+$ helicity amplitudes. And we have
verified  that soon above threshold,
the four helicity conserving amplitudes in (\ref{4HC-amp-list}) are much larger than
all other ones,  in both  SM and MSSM.

We have then established the so-called supersimple (sim) expressions for
the HC amplitudes in (\ref{4HC-amp-list}), both in SM and in MSSM.
These expressions (explicitly
written in Appendix A) are really simple and provide a panoramic view of
the dynamics; i.e., of the fermion, gauge and higgs exchanges,
and (in the supersymmetric part) of  the sfermion, additional higgses,
charginos and neutralinos exchanges.

Moreover, the accuracy of these sim expressions
is sufficient to allow their use in order to search
for possible new physics contributing in addition to  SM or MSSM.
In other words,  sim expressions may be used  to avoid the  enormous codes
needed when using the complete   1loop expressions.
Thus, analyses  done by only using Born terms, can  be easily upgraded
to the 1loop EW order.

In  previous work \cite{super, ttbar},
we have emphasized the peculiar simplicity arising in the MSSM case.
However in the process $e^-e^+\to  W^- W^+$,  the purely
supersymmetric contributions are rather small. So even in the purely
SM case, we get simple accurate expressions, that are valid at LHC energies.

At present there is no signal of supersymmetry at LHC. The discovery of the Higgs
boson at 125 GeV is nevertheless a source of questions about the
possibility of various kinds of New Physics effects \cite{Altarelli}.
The process $e^-e^+\to  W^- W^+$ is a typical place where such
effects can be looked for. For our illustrations,
we have taken the cases of AGC or $Z'$ contributions,
which have been often discussed. Other possibilities may of course
be tried \cite{Andreev}.

Our supersimple
expressions are intended to help  differentiating  such New Physics  effects from
standard or supersymmetric corrections, in a way which is as simple as possible,
while at the same time allowing us to directly  see  the responsible dynamics. \\

\vspace*{1cm}

\renewcommand{\thesection}{A}
\renewcommand{\theequation}{A.\arabic{equation}}
\setcounter{equation}{0}

\section{Appendix: Supersimple expressions for the 4 HC amplitudes}

The purpose of this Appendix is to present the {\it supersimple} (sim) expressions
for the four leading HC amplitudes listed in (\ref{4HC-amp-list}).
The  procedure  is valid for of any 2-to-2 process at 1loop EW order,
in either MSSM or SM, provided the Born contribution is non-negligible.
And it is based on the fact that the
 helicity conserving (HC) amplitudes, are the only relevant
 ones at high energy \cite{heli1,heli2}.

 To derive these sim expressions, we start from a complete 1loop EW order calculation,
 and then take the high energy
limit using \cite{asPV}. As in the analogous cases studied in \cite{super, ttbar},
these expressions
constitute a very good high energy approximation, to the HC amplitudes, renormalized
on-shell \cite{OS}.

Apart from  possible additive constants,
these sim expressions consist of linear combinations of just four forms  \cite{super, ttbar}.
For   $e^-e^+\to W^-W^+$, the structure of these forms     simplifies   as
\bqa
&& \overline{\ln^2x_{Vi}}  =  \ln^2x_{V}+4L_{aVi} ~~,~~
x_V \equiv \left (\frac{-x-i\epsilon}{m_V^2} \right )~~,  \label{Sud-ln2-form} \\
&& \overline{\ln x_{ij}}  =  \ln x_{ij}+b^{ij}_0(m_a^2)-2 ~~ , ~~
\ln x_{ij}\equiv \ln \frac{-x-i\epsilon}{m_im_j} ~~, \label{Sud-ln-form} \\
&& \overline{\ln^2r_{xy}}=\ln^2r_{xy}+ \pi^2 ~~~,~~~
r_{xy} \equiv \frac{-x-i\epsilon}{-y-i\epsilon} ~~~~, \label{ln2r-form} \\
&& \ln r_{xy} ~~~,~~ \label{lnr-form}
\eqa
where $(x,y)$ denotes any two of the Mandelstam variables $(s,t,u)$.

The indices $(i,j,V)$ in the first two forms  (\ref{Sud-ln2-form}, \ref{Sud-ln-form}),
called Sudakov augmented forms \cite{super},  denote internally exchanged particles,
in the various 1loop diagrams; while  $V$ always
refers to a gauge exchange. The index "$a$" always refers to a  particle such that
the tree-level vertices $aVi$ or $aij$ are non-vanishing. This particle $a$, could
either be an external particle (i.e. $e^\mp$ or $W^\mp$ for the process studied here),
or a particle  contributing  at tree level through an exchange in the $s,~t$ or $u$
channel (i.e. $\nu_e$, or\footnote{As always, for  an internal photon we use a mass $m_\gamma$,
in order to  regularize possible infrared singularities.} $\gamma, Z$ in our case).
Using these, the energy-independent
expressions in (\ref{Sud-ln2-form}, \ref{Sud-ln-form}) may be written as
\cite{super, ttbar, asPV}
\bqa
 L_{aVi}& = & \rm Li_2 \left ( \frac{2m_a^2+i\epsilon}{m_V^2-m_i^2+m_a^2+i\epsilon +
\sqrt{\lambda (m_a^2+i\epsilon, m_V^2, m_i^2)}} \right )
\nonumber \\
&& + \rm Li_2 \left ( \frac{2m_a^2+i\epsilon }{m_V^2-m_i^2+m_a^2+i\epsilon -
\sqrt{\lambda (m_a^2+i\epsilon, m_V^2, m_i^2)}} \right )~~,\label{LaVi-term} \\[0.5cm]
b_0^{ij}(m_a^2)& \equiv& b_0(m_a^2; m_i,m_j) =
2 + \frac{1}{m_a^2} \Big [ (m_j^2 -m_i^2)\ln\frac{m_i}{m_j}\nonumber\\
&& + \sqrt{\lambda(m_a^2+i\epsilon, m_i^2, m_j^2)}  {\rm ArcCosh} \Big
(\frac{m_i^2+m_j^2-m_a^2-i\epsilon}{2 m_i m_j} \Big ) \Big ] ~~, \label{b0ij}
\eqa
where
\bq
\lambda(a,b,c)=a^2+b^2+c^2-2ab-2ac-2bc~~. \label{lambda-function}
\eq

The other two forms (\ref{ln2r-form}, \ref{lnr-form}) are solely induced
  by box contributions to the  asymptotic amplitudes \cite{asPV}.
  The forms (\ref{lnr-form}) in particular, have no dependence on mass scales and  never  arise
 from  differences of the augmented  Sudakov linear-log  contributions,
 of the type  (\ref{Sud-ln-form}).\\

As already said, apart from  possible additive constants,
 the sim expressions consist of linear combinations of  the
 four forms (\ref{Sud-ln2-form}-\ref{lnr-form}).
The coefficients of these forms  may
 involve ratios of Mandelstam variables, as well as constants.
Particularly for  the Sudakov augmented forms
(\ref{Sud-ln2-form}, \ref{Sud-ln-form}) though, their coefficients should be
such that, when differences in the  scales of masses  and   Mandelstam variables are disregarded,
 then, the complete  coefficients in the implied e.g.  $\ln s $  and $\ln^2 s$ terms
become the constants given by  general   rules
\cite{MSSMrules1,MSSMrules2,MSSMrules3,MSSMrules4}.

Generally, these  \emph{supersimple} HC helicity amplitudes, differ from the
on-shell renormalized ones \cite{OS}, by   small additive constant terms,
in  both, the MSSM and SM cases.
We have checked numerically that
for the  process studied here, these  are indeed negligible.

In the next two subsections we give the \emph{supersimple} expressions
for the $e^-e^+\to W^-_TW^+_T$ and $e^-e^+\to W^-_LW^+_L$  HC amplitudes respectively.
In these, we first give the results for the case where infrared singularities are
regularized by using $m_\gamma=m_Z$ \cite{super, ttbar};
and subsequently quote the corrections
for the  $m_\gamma  \neq m_Z$  case.
In each case, we give separately the SM and the MSSM predictions.

\subsection{The $e^-e^+\to W^-_TW^+_T$ HC amplitudes}

There are  two such  HC amplitudes listed in the left part of
(\ref{4HC-amp-list}); namely $F_{-{1\over2}-+}$ and $F_{-{1\over2}+-}$.
In the $m_\gamma=m_Z$ case,  using the  Born results
in (\ref{Born-asym-TT-HC}),

\vspace*{0.1cm}

\noindent
the  asymptotic supersimple {\bf sim SM } expressions  are
\bqa
F_{-{1\over2}-+}&=&F^{\rm Born}_{-{1\over2}-+} \left ({\alpha\over16\pi s^2_W}\right )
\Bigg \{\overline{\ln t_{Ze}}\left ({3+2c^2_W\over c^2_W}-{4t\over s}+{4s\over u}\right )
+\overline{\ln t_{W\nu}}\left ({-1+10c^2_W\over c^2_W}-{8t\over s} \right )\nonumber\\
&&
+{\overline{\ln t_{Z\nu}}\over c^2_W} +2\overline{\ln t_{We}}
+\overline{\ln u_{Ze}}\left ({4t\over u}-{4t\over s} \right )
+{8t\over s}(\overline{\ln s_{W\nu}}+\overline{\ln s_{Ze}})-4\overline{\ln u_{W\nu}}
\nonumber\\
&&-3\overline{\ln^2t_{Ze}}-\overline{\ln^2t_{ZW}}
-3\overline{\ln^2t_{W\nu}}-\overline{\ln^2t_{WZ}}\nonumber\\
&&- {1\over c^2_W}(\overline{\ln^2s_{Ze}}
+4c^2_W\overline{\ln^2s_{ZW}})
-2\overline{\ln^2s_{WZ}}+2\overline{\ln^2u_{Ze}}+2\overline{\ln^2u_{ZW}}
\nonumber\\
&&-{2t\over u}(\overline{\ln^2s_{W\nu}}+\overline{\ln^2s_{WZ}}-\overline{\ln^2t_{W\nu}}
-\overline{\ln^2t_{WZ}})\nonumber\\
&&+\overline{\ln^2r_{ts}}\left [{2u^3+2t^3+6ut^2+2tu^2)\over 2u^3c^2_W}
+{6u^3-6t^3)\over u^3} \right ]\nonumber\\
&&+{4s\over u}\overline{\ln^2r_{ut}}+{4(t-u)\over u}\overline{\ln^2r_{us}}
+\left [{t(2t+5u)\over u^2c^2_W}+{t(12t^2+6u^2+6tu)\over su^2}\right ]\ln r_{ts}
\nonumber\\
&&
+~{t(16u+12t)\over su}\ln r_{us}-\left ({8t\over u}+4 \right)\ln r_{tu}
+~{t(1-6c^2_W)\over uc^2_W} \Bigg\} ~~,
\label{simSM--+} \\[0.5cm]
F_{-{1\over2}+-}&=&F^{\rm Born}_{-{1\over2}+-}\left ({\alpha\over16\pi s^2_W} \right )
\Bigg\{\overline{\ln t_{Ze}}\left ({3+2c^2_W\over c^2_W}- {4t\over s}+{4s\over u}\right )
+\overline{\ln t_{W\nu}}\left ({-1+10c^2_W\over c^2_W}-{8t\over s}\right )\nonumber\\
&&
+{1\over c^2_W}\overline{\ln t_{Z\nu}} +2\overline{\ln t_{We}}
+\overline{\ln u_{Ze}}\left ({4t\over u}-{4t\over s} \right )
+ {8t\over s} \left (\overline{\ln s_{W\nu}}+\overline{\ln s_{Ze}}\right )
-4\overline{\ln u_{W\nu}}
\nonumber\\
&&-3\overline{\ln^2t_{Ze}}-\overline{\ln^2t_{ZW}}
-3\overline{\ln^2t_{W\nu}}-\overline{\ln^2t_{WZ}}\nonumber\\
&&- {1\over c^2_W} (\overline{\ln^2s_{Ze}} +4c^2_W\overline{\ln^2s_{ZW}})
-2\overline{\ln^2s_{WZ}}+2\overline{\ln^2u_{Ze}}+2\overline{\ln^2u_{ZW}}
\nonumber\\
&&- {2t\over u}\left (\overline{\ln^2s_{W\nu}}+\overline{\ln^2s_{WZ}}
-\overline{\ln^2t_{W\nu}}-\overline{\ln^2t_{WZ}} \right )
+\overline{\ln^2r_{ts}}\left [{u-t\over uc^2_W}+{6(u-t)\over u} \right ]\nonumber\\
&&
+{4s\over u}\overline{\ln^2r_{ut}}
+({4t^2+2ut+6u^2\over ut})\overline{\ln^2r_{us}}
+\left [{-3\over c^2_W}+{18u^2+30ut\over su} \right ]
\ln r_{ts}\nonumber\\
&&
+({4t\over u}+8)\ln r_{tu}+ ({4t\over s}+12)\ln r_{us}
- {1-6c^2_W\over c^2_W}\Bigg\} ~~, \label{simSM-+-}
\eqa

\vspace*{0.5cm}

\noindent
while the   {\bf sim  MSSM}  results, always assuming CP conservation,
are
\bqa
F_{-{1\over2}-+}&=&F^{\rm Born}_{-{1\over2}-+}\left ({\alpha\over16\pi s^2_W} \right )
\Bigg\{{1\over c^2_W}\left ( 3\overline{\ln t_{Ze}}
-\overline{\ln t_{W\nu}}+\overline{\ln t_{Z\nu}} \right ) -2\overline{\ln t_{Ze}}
\nonumber\\
&& +6\overline{\ln t_{W\nu}}
+2\overline{\ln t_{We}}-3\overline{\ln^2t_{Ze}}-\overline{\ln^2t_{ZW}}
-3\overline{\ln^2t_{W\nu}}-\overline{\ln^2t_{WZ}}\nonumber\\
&&- {1\over c^2_W}(\overline{\ln^2s_{Ze}} +4c^2_W\overline{\ln^2s_{ZW}})
-2\overline{\ln^2s_{WZ}}+2\overline{\ln^2u_{Ze}}+2\overline{\ln^2u_{ZW}}
\nonumber\\
&&-~{2t\over u}(\overline{\ln^2s_{W\nu}}+\overline{\ln^2s_{WZ}}
-\overline{\ln^2t_{W\nu}}-\overline{\ln^2t_{WZ}})
\nonumber\\
&&+{4s\over u}\ln r_{tu}-{12t^2\over su}\ln r_{ts}+({4t\over s}-{8t\over u})\ln r_{us}
+{2t\over uc^2_W}\ln r_{ts}
\nonumber\\
&&-~{1\over c^2_W}
\Big \{\sum_j|Z^N_{1j}s_W+Z^N_{2j}c_W|^2\overline{\ln t_{\chi^0_j\tilde{e_L}}}
+2c^2_W\sum_j|Z^+_{1j}|^2\overline{\ln t_{\chi^+_j\tilde{\nu}}} \Big \}
\nonumber\\
&&+\overline{\ln^2r_{ts}}\left [{t^2+u^2\over u^2c^2_W}+{6t^2+6u^2)\over u^2}\right ]
+~{4s\over u}\overline{\ln^2r_{ut}}+{4(t-u)\over u}\overline{\ln^2r_{us}}\Bigg\} ~~,
\label{simMSSM--+} \\[0.5cm]
F_{-{1\over2}+-}&=&F^{\rm Born}_{-{1\over2}+-}\left ({\alpha\over16\pi s^2_W} \right )
\Bigg\{{1\over c^2_W}[3\overline{\ln t_{Ze}}
-\overline{\ln t_{W\nu}}+\overline{\ln t_{Z\nu}}] -2\overline{\ln t_{Ze}}
\nonumber\\
&& +6\overline{\ln t_{W\nu}} +2\overline{\ln t_{We}}
-3\overline{\ln^2t_{Ze}}-\overline{\ln^2t_{ZW}}
-3\overline{\ln^2t_{W\nu}}-\overline{\ln^2t_{WZ}}
\nonumber\\
&&-~{1\over c^2_W}(\overline{\ln^2s_{Ze}}
+4c^2_W\overline{\ln^2s_{ZW}})-2\overline{\ln^2s_{WZ}}
+2\overline{\ln^2u_{Ze}}+2\overline{\ln^2u_{ZW}}
\nonumber\\
&&-~{2t\over u}(\overline{\ln^2s_{W\nu}}+\overline{\ln^2s_{WZ}}
-\overline{\ln^2t_{W\nu}}-\overline{\ln^2t_{WZ}})
\nonumber\\
&&+~{12(t-s)\over s}\ln r_{ts}+({4t\over s}+8)\ln r_{us}
-~({2\over c^2_W})\ln r_{ts}-{4s\over u}\ln r_{tu}
\nonumber\\
&&-~{1\over c^2_W}
\Big \{\sum_j|Z^N_{1j}s_W+Z^N_{2j}c_W|^2\overline{\ln t_{\chi^0_j\tilde{e_L}}}
+2c^2_W\sum_j|Z^+_{1j}|^2\overline{\ln t_{\chi^+_j\tilde{\nu}}} \Big \}
\nonumber\\
&&+\overline{\ln^2r_{ts}}[{u-t\over uc^2_W}+{6(u-t)\over u}]
+{4s\over u}\overline{\ln^2r_{ut}}+{4(t^2+u^2)\over ut}\overline{\ln^2r_{us}}
\Bigg\} ~~, \label{simMSSM-+-}
\eqa
where  the indices $(i,j)$ in (\ref{simMSSM--+}, \ref{simMSSM-+-}) and  (\ref{simMSSM+00}, \ref{simMSSM-00}) below,  refer to chargino and neutralino contributions, defined as in  \cite{Rosiek}.

Note  the constant terms at the end of the r.h.s. of the SM results
(\ref{simSM--+}, \ref{simSM-+-}).
No such constants appear in the corresponding  MSSM amplitudes
(\ref{simMSSM--+}, \ref{simMSSM-+-}).\\

 In the  $m_\gamma  \neq m_Z$ case, the correction to be added to
  (\ref{simSM--+}-\ref{simMSSM-+-}),
is  given by
\bqa
&& \delta F_{-{1\over2}\mp \pm}= F^{\rm Born}_{-{1\over2}\mp \pm}
\left ({\alpha\over16\pi s^2_W}\right )
\Bigg [ \Bigg \{-2s^2_W( \overline{\ln^2t_{\gamma e}}+\overline{\ln^2t_{\gamma W}})
+16 s^2_W {t\over s} \overline{\ln s_{\gamma e}}\nonumber\\
&&+2s^2_W[-2\overline{\ln^2s_{\gamma e}}+8\overline{\ln t_{\gamma e}}]
 -2s^2_W[2\overline{\ln^2s_{\gamma W}}+\overline{\ln^2 t_{W\gamma}}]\nonumber\\
&&+2s^2_W[-2\overline{\ln^2s_{W\gamma}}-\overline{\ln^2 t_{\gamma e}}
-\overline{\ln^2 t_{\gamma W}}+4(1-{t\over s})\overline{\ln t_{\gamma e}}]\nonumber\\
&&+2s^2_W \Big [-2{t\over u}\overline{\ln^2s_{W\gamma}}-{t\over u}
(\overline{\ln^2 u_{\gamma e}}
+\overline{\ln^2 u_{\gamma W}})+4({t\over u}-{t\over s})\overline{\ln u_{\gamma e}}\Big ]
\nonumber\\
&&-2s^2_W \Big [{s-u\over u}(\overline{\ln^2u_{\gamma e}}+\overline{\ln^2 u_{\gamma W}})
+{s-t\over u}\overline{\ln^2 t_{W\gamma}}+4(2+{t\over u})\overline{\ln t_{\gamma e}} \Big ]
\Bigg\}
\nonumber \\
&& -\Big \{ m_\gamma \to m_Z       \Big \} \Bigg ]~~, \label{deltaFTT}
\eqa
where (\ref{Born-asym-TT-HC})  is again  used.

\vspace*{1cm}
\subsection{The $e^-e^+\to W^-_L W^+_L$ HC amplitudes}

In the $m_\gamma=m_Z$ case, using the asymptotic Born LL amplitudes
(\ref{Born-asym-LL-HC}), \\
\noindent
the high energy supersimple  {\bf  sim SM}  results  are  written as\\
\bqa
F_{+{1\over2}00}&=&F^{\rm Born}_{+{1\over2}00}\Bigg\{\left ({\alpha\over4\pi} \right )
\Big \{{1\over c^2_W}
\left [-\overline{\ln^2s_{Ze}}+3\overline{\ln s_{Ze}}-1 \right ]
+{1\over 4s^2_Wc^2_W}\left [-\overline{\ln^2s_{ZW}}+4\overline{\ln s_{ZW}} \right ]
\nonumber\\
&&+{1\over 2s^2_W}\left [-{1\over2}(\overline{\ln^2s_{WZ}}+\overline{\ln^2s_{WH_{SM}}})
+2\overline{\ln s_{WZ}} +2\overline{\ln s_{WH_{SM}}}\right ]
\nonumber\\
&&-{3(m^2_t+m^2_b)\over 2s^2_Wm^2_W}\overline{\ln s_{tb}}
-{1\over 4c^2_W} \Big [4(\overline{\ln^2t_{ZW}}-\overline{\ln^2u_{ZW}})
+{2(u-t)\over u}\overline{\ln^2r_{ts}}
\nonumber\\
&& -{2(t-u)\over t}\overline{\ln^2r_{us}} \Big ]\Big \}
+\Sigma^{\rm seSM}\left (+{1\over2},0,0 \right )\Bigg\}~~,
\label{simSM+00} \\[0.5cm]
F_{-{1\over2}00}&=&F^{\rm Born}_{-{1\over2}00}\Bigg\{\left ({\alpha\over4\pi} \right)
\Big \{{1\over 4s^2_Wc^2_W}
\left[-\overline{\ln^2s_{Ze}}+3\overline{\ln s_{Ze}}-1 \right ]
\nonumber\\
&&-{(1-2s^2_W)\over 2s^2_W}[-\overline{\ln^2s_{W\nu}}+3\overline{\ln s_{W\nu}}-1]
\nonumber\\
&& +{2c^2_W\over s^2_W}\left [{1\over2}\overline{\ln s_{W\nu}}+{1\over2}
+2\overline{\ln s_{WW}}\right ]
+{1\over 4s^2_Wc^2_W}\left [-\overline{\ln^2s_{ZW}}+4\overline{\ln s_{ZW}} \right ]
\nonumber\\
&&+{(1-2c^2_W)\over 2s^2_W}
\left [-{1\over2}(\overline{\ln^2s_{WZ}}+\overline{\ln^2s_{WH_{SM}}})+2\overline{\ln s_{WZ}}
+2\overline{\ln s_{WH_{SM}}} \right ]
\nonumber\\
&&+{c^2_W\over s^2_W}\left [\overline{\ln s_{WZ}}
+\overline{\ln s_{WH_{SM}}}\right ]-{3(m^2_t+m^2_b)\over 2s^2_Wm^2_W}\overline{\ln s_{tb}}
\nonumber\\
&&-{c^2_W\over 4s^2_W} \left [4\overline{\ln^2t_{W\nu}}
+2\overline{\ln^2t_{WZ}}+2\overline{\ln^2t_{WH_{SM}}}
-4(1-{t\over u})\overline{\ln^2r_{ts}}\right ]
\nonumber\\
&&-{1\over 8c^2_Ws^2_W}\left [4(\overline{\ln^2t_{ZW}}-\overline{\ln^2u_{ZW}})
+{2(u-t)\over u}\overline{\ln^2r_{ts}}
-{2(t-u)\over t}\overline{\ln^2r_{us}} \right ]\Big \}
\nonumber\\
&& +\Sigma^{\rm seSM}\left (-{1\over2},0,0 \right )\Bigg\}~~, \label{simSM-00}
\eqa

\vspace*{0.5cm}

\noindent
while the supersimple  {\bf sim  MSSM}  results  are
\bqa
F_{+{1\over2}00}&=&F^{\rm Born}_{+{1\over2}00}
\Bigg\{\left ({\alpha\over4\pi} \right ) \Big \{{1\over c^2_W}
 \left [-\overline{\ln^2s_{Ze}}+3\overline{\ln s_{Ze}}
-\Sigma_i|Z^N_{1i}|^2 \overline{\ln s_{\chi^0_i\tilde{e_R}}} \right ]
\nonumber\\
&&+{1\over 4s^2_Wc^2_W}\left [-\overline{\ln^2s_{ZW}}+4\overline{\ln s_{ZW}} \right ]
+{1\over 2s^2_W}\left [-{1\over2} \overline{\ln^2s_{WZ}} +2\overline{\ln s_{WZ}} \right ]
\nonumber\\
&&-{1\over 4s^2_W} \left [\cos^2(\beta-\alpha)\overline{\ln^2 s_{WH^0}}+\sin^2(\beta-\alpha)
\overline{\ln^2 s_{Wh^0}} \right ]
\nonumber\\
&&+{1\over 2s^2_W} \left [2\cos^2(\beta-\alpha)\overline{\ln s_{WH^0}}
+2\sin^2(\beta-\alpha)\overline{\ln s_{Wh^0}} \right ]
\nonumber\\
&&-{1\over 2s^2_Wc^2_W}\Sigma_{ij}
\Big [\Big |{1\over\sqrt{2}}Z^-_{2i}(Z^N_{1j}s_W+Z^N_{2j}c_W)-Z^-_{1i}Z^N_{3j}c_W \Big |^2
\nonumber\\
&& +\Big |{1\over\sqrt{2}}Z^+_{2i}(Z^N_{1j}s_W+Z^N_{2j}c_W)
+Z^+_{1i}Z^N_{4j}c_W \Big |^2  \Big ] \overline{\ln s_{\chi^+_i\chi^0_j}}
\nonumber\\
&&-{3(m^2_t+m^2_b)\over 2s^2_Wm^2_W}\overline{\ln s_{tb}}
-{\cos^2\beta\over2c^2_W} \left [{s\over u}\overline{\ln^2r_{ts}}-
{s\over t}\overline{\ln^2r_{us}} \right ]
\nonumber\\
&& -{1\over 4c^2_W}\left [4(\overline{\ln^2t_{ZW}}-\overline{\ln^2u_{ZW}})
+{2(u-t)\over u}\overline{\ln^2r_{ts}}
-{2(t-u)\over t}\overline{\ln^2r_{us}} \right ] \Big \}
\nonumber\\
&& +\Sigma^{\rm seMSSM}\left (+{1\over2},0,0 \right )\Bigg\} ~~,
\label{simMSSM+00} \\[0.5cm]
F_{-{1\over2}00}&=&F^{\rm Born}_{-{1\over2}00}\Bigg\{\left ({\alpha\over4\pi} \right )
\Big \{{1\over 4s^2_Wc^2_W} \Big [-\overline{\ln^2s_{Ze}}+3\overline{\ln s_{Ze}}
-\overline{\ln^2s_{ZW}}+4\overline{\ln s_{ZW}}
\nonumber\\
&& -\Sigma_i|Z^N_{1i}s_W+Z^N_{2i}c_W|^2\overline{\ln s_{\chi^0_i\tilde{e_L}}} \Big ]
\nonumber\\
&&-{(1-2s^2_W)\over 2s^2_W}\left [-\overline{\ln^2s_{W\nu}}+3\overline{\ln s_{W\nu}}
- {1\over2}\overline{\ln^2s_{WZ}}+2\overline{\ln s_{WZ}}-\Sigma_i|Z^+_{1i}|^2
\overline{\ln s_{\chi^+_i\tilde{\nu_L}}}\right ]
\nonumber\\
&&+{c^2_W\over s^2_W} \left [{\overline{\ln s_{W\nu}}}
+4\overline{\ln s_{WW}}-\Sigma_i|Z^+_{1i}|^2\overline{\ln s_{\chi^+_i\tilde{\nu_L}}} \right ]
\nonumber\\
&& -{(1-2c^2_W)\over 4s^2_W} \left [\cos^2(\beta-\alpha)\overline{\ln^2 s_{WH^0}}
+\sin^2(\beta-\alpha) \overline{\ln^2 s_{Wh^0}} \right ]
\nonumber\\
&&+{(1-2c^2_W)\over s^2_W} \left [ \cos^2(\beta-\alpha)\overline{\ln s_{WH^0}}
+\sin^2(\beta-\alpha)\overline{\ln s_{Wh^0}} \right ]
\nonumber\\
&&+{c^2_W\over s^2_W} \left [\overline{\ln s_{WZ}}
+\cos^2(\beta-\alpha) \overline{\ln s_{WH^0}}
+\sin^2(\beta-\alpha) \overline{\ln s_{Wh^0}}\right ]
\nonumber\\
&&-{3(m^2_t+m^2_b)\over 2s^2_Wm^2_W}\overline{\ln s_{tb}}-{1\over 2s^2_Wc^2_W}\Sigma_{ij}
\Big [\Big |{1\over\sqrt{2}}Z^-_{2i}(Z^N_{1j}s_W+Z^N_{2j}c_W)-Z^-_{1i}Z^N_{3j}c_W \Big |^2
\nonumber\\
&&+\Big |{1\over\sqrt{2}}Z^+_{2i}(Z^N_{1j}s_W+Z^N_{2j}c_W)
+Z^+_{1i}Z^N_{4j}c_W \Big |^2  \Big ] \overline{\ln s_{\chi^+_i\chi^0_j}}
\nonumber\\
&&- {c^2_W\over 4s^2_W} \left [4\overline{\ln^2t_{W\nu}}
+2\overline{\ln^2t_{WZ}}
-4\left (1-{t\over u} \right )\overline{\ln^2r_{ts}} \right ]
\nonumber\\
&&- {c^2_W\over 2s^2_W}
\left [\cos^2(\beta-\alpha)\overline{\ln^2 t_{WH^0}}+\sin^2(\beta-\alpha)
\overline{\ln^2 t_{Wh^0}} \right ]
\nonumber\\
&&-{1\over 8c^2_Ws^2_W}\left [4(\overline{\ln^2t_{ZW}}-\overline{\ln^2u_{ZW}})
+{2(u-t)\over u}\overline{\ln^2r_{ts}}
-{2(t-u)\over t}\overline{ln^2r_{us}} \right ]
\nonumber\\
&&-{\sin^2\beta\over2c^2_Ws^2_W} \left [{s\over u}\overline{\ln^2r_{ts}}-
{s\over t}\overline{\ln^2r_{us}} \right ]
-{c^2_W \sin^2\beta \over s^2_W} {s\over u}\overline{\ln^2r_{ts}} \Big \}
\nonumber\\
&& +\Sigma^{\rm seMSSM}\left (-{1\over2},0,0 \right )\Bigg\}~~. \label{simMSSM-00}
\eqa\\

 In the  $m_\gamma  \neq m_Z$ case,
 the correction  to be added to (\ref{simSM+00}-\ref{simMSSM-00}) is     given by
 \bqa
 \delta F_{\pm {1\over2} 00}&= &F^{\rm Born}_{\pm{1\over2}00}\left ({\alpha\over4\pi} \right )
\Bigg [ \Big \{ -\overline{\ln^2s_{\gamma e}}+3\overline{\ln s_{\gamma e}}
 -\overline{\ln^2s_{\gamma W}}
 \nonumber\\
&& +4\overline{\ln s_{\gamma W}}
-2 \overline{\ln^2t_{\gamma W}} +2 \overline{\ln^2u_{\gamma W}} \Big \}
- \Big \{    m_\gamma \to m_Z    \Big \} \Bigg ]~~ , \label{deltaF00}
\eqa
where   (\ref{Born-asym-LL-HC}) is again used.

The $\Sigma^{\rm se}$-contributions in either (\ref{simSM+00}-\ref{simSM-00})
or (\ref{simMSSM+00}-\ref{simMSSM-00}), respectively appearing  in SM and MSSM,
 come from  the photon and Z self-energy contributions together with
 their renormalization counter terms. Their
 explicit expressions are
\bqa
\Sigma^{\rm se}\left (-{1\over2},0,0 \right )&= &
{-4s^2_Wc^2_W\over s}\left \{\hat{\Sigma}_{\gamma\gamma}(s)
+{1-2s^2_W\over s_Wc_W}\hat{\Sigma}_{Z\gamma}(s)+{(1-2s^2_W)^2\over4s^2_Wc^2_W}
\hat{\Sigma}_{ZZ}(s) \right \} \nonumber \\
&  + & C_P ~~, \label{Cse-00} \\[0.5cm]
\Sigma^{\rm se}\left (+{1\over2},0,0 \right )&=&
{-2c^2_W\over s} \left \{\hat{\Sigma}_{\gamma\gamma}(s)
+{1-4s^2_W\over 2s_Wc_W}\hat{\Sigma}_{Z\gamma}(s)-{(1-2s^2_W)\over2c^2_W}
\hat{\Sigma}_{ZZ}(s)\right \}~~, \label{Cse+00}
\eqa
where the renormalized gauge self energies   $\hat{\Sigma}$ can be found in
\cite{ttbar}, together with their supersimple approximations. The last term in
(\ref{Cse-00}), given by
\bq
C_P=-{\alpha c^2_W\over\pi s^2_W}\overline{\ln s_{WW}}~~, \label{pinch-part}
\eq
comes from  the pinch part  that  had been previously
removed from the left and right triangular  contributions, and is here restored
\cite{pinch,pinch1}.

Note that no such $\Sigma^{\rm se}$-contributions exist for the transverse amplitudes
 in (\ref{simSM--+}-\ref{simMSSM-+-}).

As it should,  the high energy  ln and ln-squared parts  of all expressions
(\ref{simSM--+}- \ref{simMSSM-00}),  agree with the usual Sudakov
rules and  the renormalization group results
\bqa
&& A^{RG}=-{ln\over 4\pi^2} (g^4\beta{dA^{Born}\over dg^2}
+g^{-4}\beta'{dA^{Born}\over dg^{'2}}) ~~, \nonumber \\
&& \beta^{SM}={43\over24}-{N_f\over3}~~,~~\beta^{SUSY}=-{13\over24}-{N_f\over6}
~~,~~N_f=3  ~~, \nonumber \\
&& \beta^{'SM}=-{1\over24}-{5N_f\over9}~~,~~\beta^{'SUSY}=-~{5\over24}-{5N_f\over18}
~~, \label{renorm-group}
\eqa
discussed in  \cite{MSSMrules1,MSSMrules2,MSSMrules3,MSSMrules4}.

\vspace*{1cm}

\renewcommand{\thesection}{B}
\renewcommand{\theequation}{B.\arabic{equation}}
\setcounter{equation}{0}

\section{ Appendix:  AGC and $Z'$ amplitudes}

\subsection{The  AGC amplitudes}

As an  Anomalous Gauge Coupling (AGC) model induced by
s-channel $\gamma $  and $Z$ exchanges with
 5 anomalous couplings $\delta_Z$, $x_{\gamma,Z}$, $y_{\gamma,Z}$, we consider
 the one presented in \cite{heliWW} and Table V of \cite{Andreev}.
In terms of these couplings and the SM ones in (\ref{e-couplings}),
the induced AGC  contributions to the TT, TL, LT and LL amplitudes,
to lowest order, are\footnote{Compare with
(\ref{FBorn-TT}, \ref{FBorn-TL}, \ref{FBorn-LT},\ref{FBorn-LL}).}
\bqa
F^{\rm AGC}_{\lambda\mu\mu}(\theta)&=& {(2\lambda)se^2\over8}(1+\mu\mu')\beta_W\sin\theta
\Bigg \{ {\delta_Z(a_{eL}\delta_{\lambda,-}+a_{eR}\delta_{\lambda,+})\over
s^2_W(s-m^2_Z)}
\nonumber\\
&&-\left [{y_{\gamma}\over s}-{y_{Z}(a_{eL}\delta_{\lambda,-}+a_{eR}\delta_{\lambda,+})
\over 2s^2_W(s-m^2_Z)} \right ]{s\over m^2_W} \Bigg \} ~~, \label{FAGC-TT} \\[0.5cm]
F^{\rm AGC}_{\lambda \mu 0}(\theta)&=&
- {(2\lambda)s\beta_W\sqrt{s}e^2\over4\sqrt{2}m_W}(2\lambda+\mu\cos\theta)
\Bigg \{ {\delta_Z(a_{eL}\delta_{\lambda,-}+a_{eR}\delta_{\lambda,+})\over
s^2_W(s-m^2_Z)}
\nonumber\\
&&-\left [{(x_{\gamma}+y_{\gamma})\over s}-{(x_{Z}+y_{Z})
(a_{eL}\delta_{\lambda,-}+a_{eR}\delta_{\lambda,+})\over
2s^2_W(s-m^2_Z)}\right ] \Bigg \}  ~~, \label{FAGC-TL}\\[0.5cm]
F^{\rm AGC}_{\lambda 0\mu'}(\theta)&=&
- {(2\lambda)s\beta_W\sqrt{s}e^2\over4\sqrt{2}m_W}(2\lambda-\mu'\cos\theta)
\Bigg \{ {\delta_Z(a_{eL}\delta_{\lambda,-}+a_{eR}\delta_{\lambda,+})\over
s^2_W(s-m^2_Z)}
\nonumber\\
&&-\left [{(x_{\gamma}+y_{\gamma})\over s}-{(x_{Z}+y_{Z})
(a_{eL}\delta_{\lambda,-}+a_{eR}\delta_{\lambda,+})\over
2s^2_W(s-m^2_Z)} \right ] \Bigg \} ~~, \label{FAGC-LT}\\[0.5cm]
F^{\rm AHC}_{\lambda 00}(\theta)&=& {(2\lambda)s^2e^2\over4m^2_W}\beta_W\sin\theta
\Bigg \{ {\delta_Z(a_{eL}\delta_{\lambda,-}+a_{eR}\delta_{\lambda,+})\over
s^2_W(s-m^2_Z)}\left (1+{s\over2m^2_W} \right )
\nonumber\\
&&-\left [{x_{\gamma}\over s}-x_{Z}{a_{eL}\delta_{\lambda,-}+a_{eR}\delta_{\lambda,+}\over
2s^2_W(s-m^2_Z)}\right ]{s\over m^2_W} \Bigg \} ~~. \label{FAGC-LL}
\eqa
Note that  $\delta_Z$ contributes to all amplitudes, except the two
TT HC ones  (because of the vanishing of the overall coefficient $(1+\mu\mu')$ in
 (\ref{FAGC-TT}) in such a  case);
 $x_{\gamma,Z}$ contribute to all TL, LT and LL amplitudes; while
 $y_{\gamma,Z}$ contribute only to the HV  TT, TL and LT amplitudes.

In the figures, and under the name AGC1,  we present illustrations
for the purely arbitrary choice
\bq
{\rm AGC1}~~~~\Rightarrow ~~~ \delta_Z=x_{\gamma}=x_Z=0.003 ~~~,~~~ y_\gamma=y_Z=0~~~.
\label{AGC1-choice}
\eq
For AGC1,  the HV TT anomalous amplitudes behave like constants at high energy;
the HC LL ones  explode  like $s/ m^2_W$; while the LT ones increase like
$\sqrt{s}/ m^2_W$.\\

In the figures we also present results for an alternative AGC2 model in which the
$s/ m^2_W$ behavior of the HC LL anomalous amplitudes is canceled by a $t$-channel
contribution; much like it is done in the Born SM case. So we construct an
 ad-hoc model with an anomalous contribution in the
t-channel which would lead to a similar cancelation.
A simple phenomenological solution is obtained by keeping only $x_{\gamma}$ and  $x_Z$
(called now $x'_{\gamma}$ and  $x'_Z$) in (\ref{FAGC-TT}-\ref{FAGC-LL}),
and adding t-channel contributions induced  by
  left- and right-handed  $We\nu$ couplings obtained from the
 initial SM one   $g_L=e/( \sqrt{2} s_W)$, through
\bqa
g^2_L & \Rightarrow & g^2_L \left (1+2s^2_W \left
[x'_{\gamma}-{2s^2_W-1\over2s_Wc_W}x'_Z \right]
\right ) ~~~,
\nonumber \\
g^2_R & \Rightarrow  &  g^2_L (2s^2_W)\left [x'_{\gamma}-{s_W\over c_W}x'_Z \right ]
~~. \label{AGC2-gLgR}
\eqa
 This does not necessarily represent
true anomalous $We\nu$ couplings; it just represents the new
contribution necessary at high energy.
For example it may come from additional neutral fermion exchanges or
from any sort of effective interaction. In the illustrations under the AGC2 name, we use
\bq
{\rm AGC2}~~~~\Rightarrow ~~~ x'_{\gamma}=x'_Z=0.03 ~~~;\label{AGC2-choice}
\eq
these values are larger than those in (\ref{AGC1-choice}),
because of the
global suppression effect following from the high energy
cancelation between t- and s-channel terms.

If one does not want to introduce an anomalous right-handed contribution
one can just keep a non vanishing $x'_{\gamma}$ only,
and add the anomalous left-handed term
\bq
g^2_L ~ \Rightarrow ~ g^2_L (1+2s^2_Wx'_{\gamma}) ~~, \label{AGC2-gL}
\eq

In any case, investigating the origin of such anomalous terms
is beyond the scope of the present work.\\

\subsection{ The $Z'$ New Physics model}

The general form of helicity amplitudes with a $Z'$ is written in Table VI of \cite{Andreev}.
The $Z'$ contributions are very similar to the SM Z ones, with specific $Z'$ mass,
width and couplings.

In general, with arbitrary $Z'$ couplings, there is an explosion of the LL, LT and
TL amplitudes at high energies.
But, it  is again  easy to get high energy cancelation in an ad-hoc manner
by just replacing the usual $Z$ contribution
involving products of couplings like $g_{Zee}g_{ZWW}$, by $Z+Z'$ exchanges using respectively
$g_{Zee}g_{ZWW}\cos^2\Phi$ for Z and
$g_{Zee}g_{ZWW}\sin^2\Phi$ for $Z'$ (with a small value of $\Phi$).
This way, the s-channel high energy contribution will be
similar to the SM $Z$ one, and will cancel with the SM t-channel contribution.
Only around the $Z'$ peak,  will the $Z'$ contribution be observable.

For the illustrations presented in the figures under the name $Z'$,
we use  $\sin\Phi=0.05$ and $m_{Z'}=3$ TeV.

\newpage

\begin{figure}[p]
%\vspace{-1cm}
\[
\epsfig{file=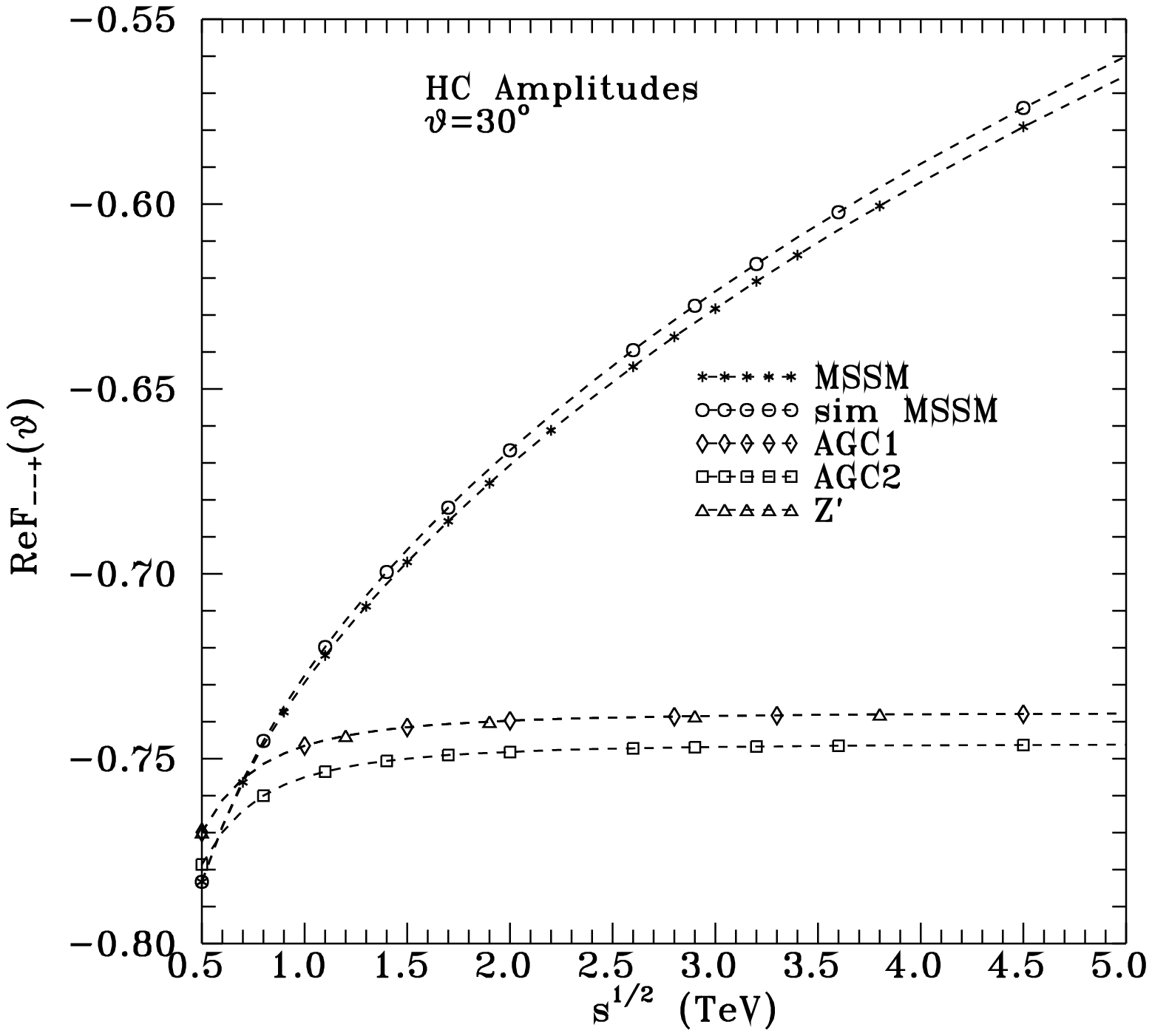, height=6.cm}\hspace{1.cm}
\epsfig{file=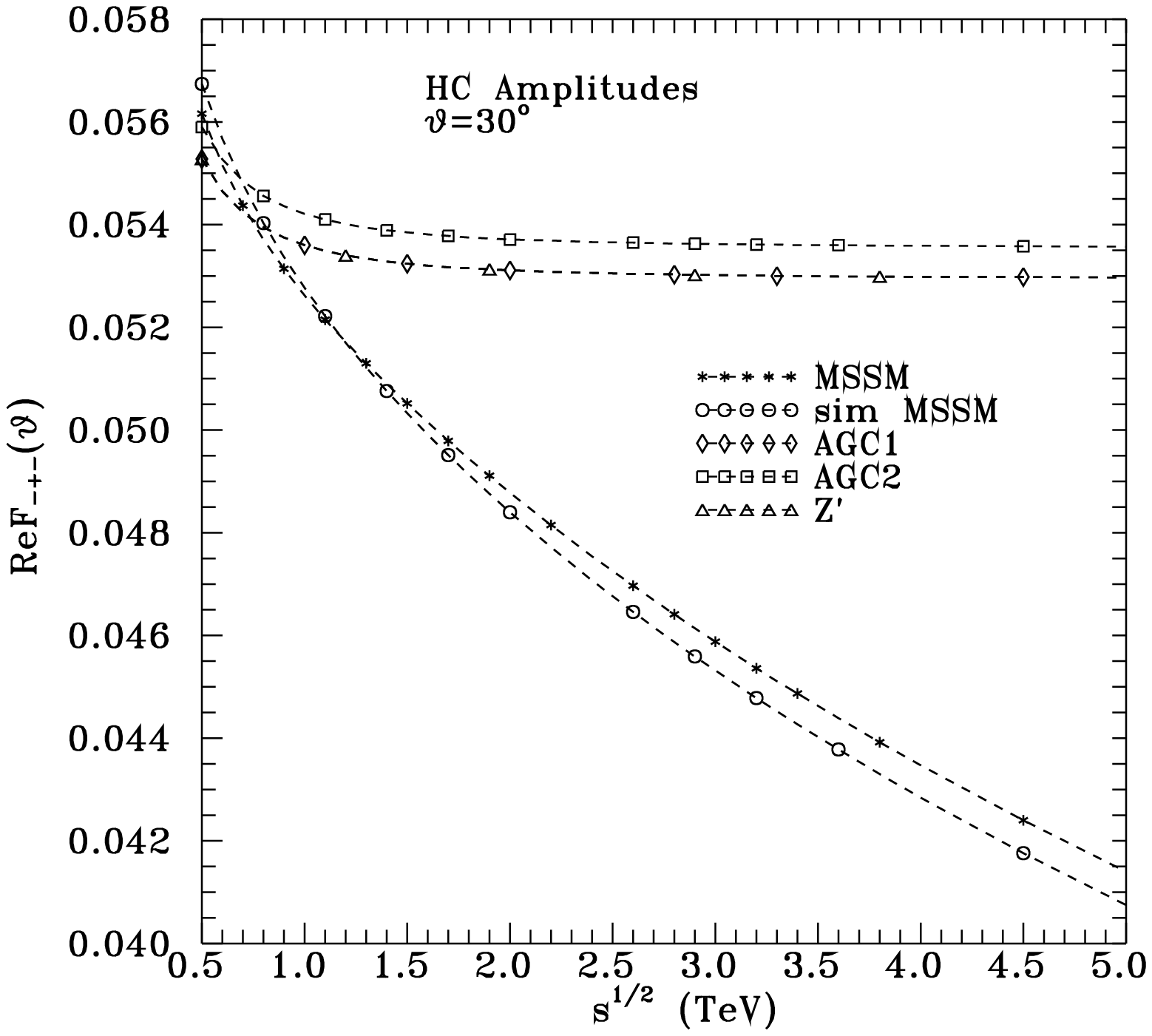,height=6.cm}
\]
\[
\epsfig{file=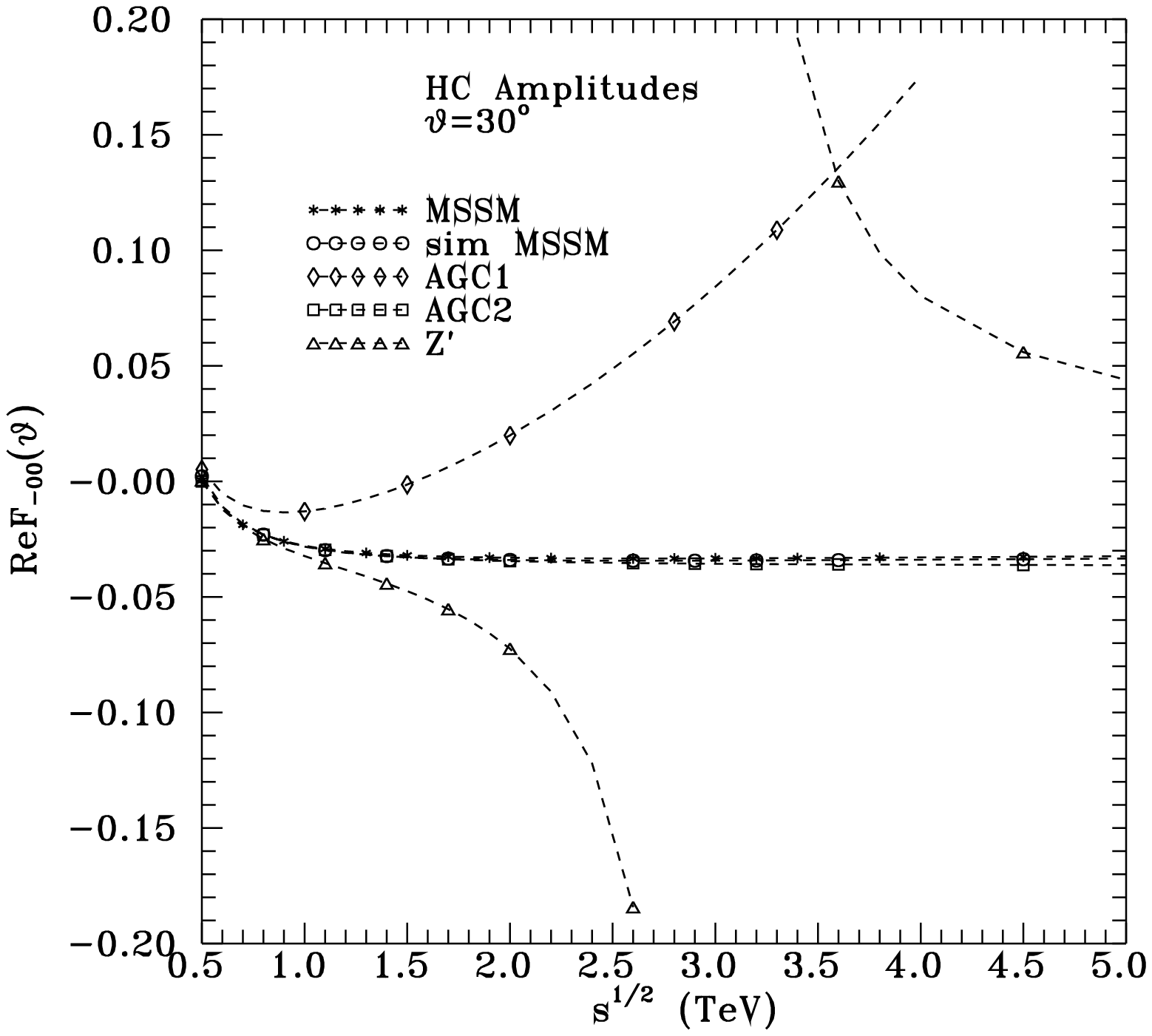, height=6.cm}\hspace{1.cm}
\epsfig{file=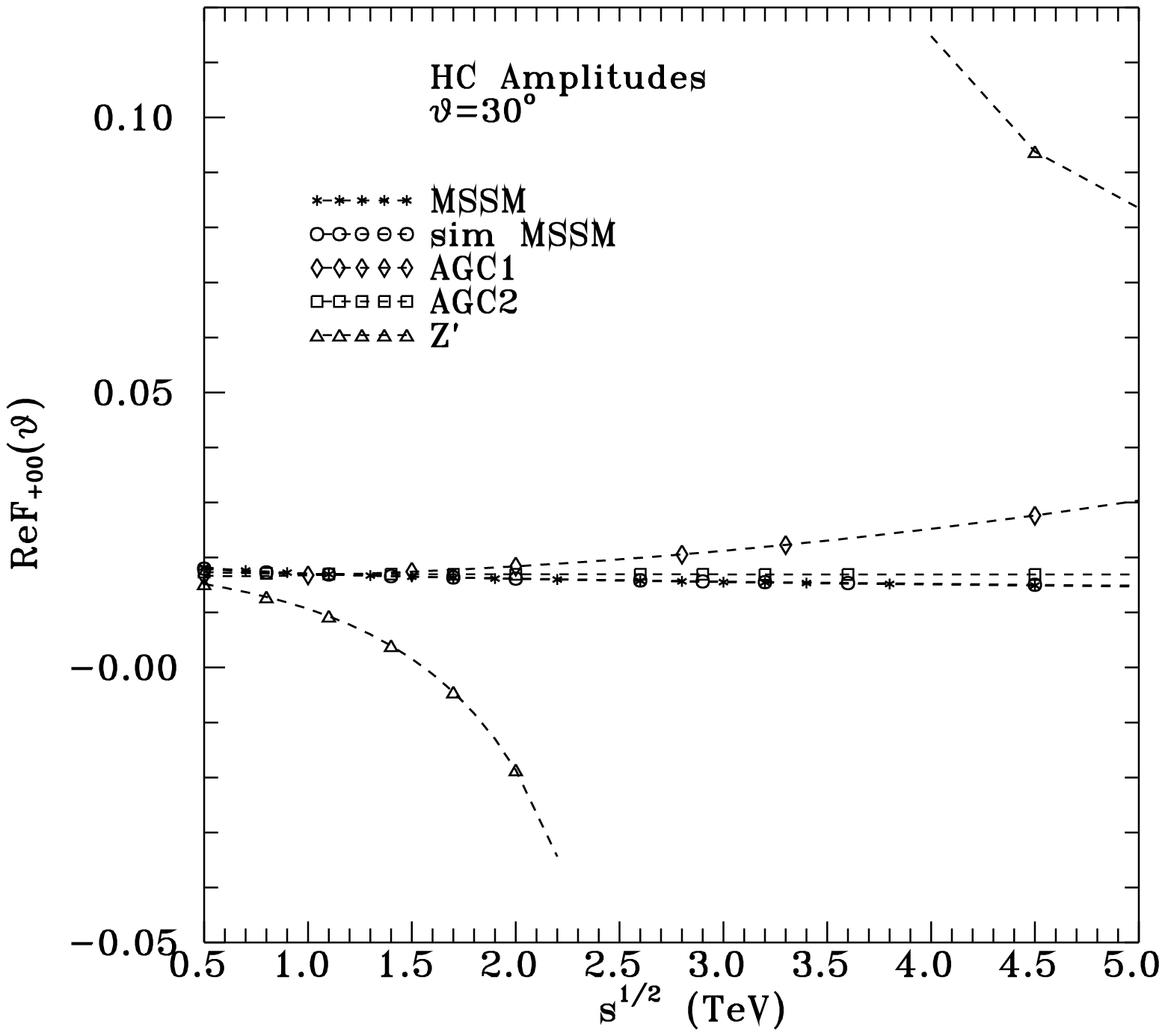,height=6.cm}
\]
\caption[1]{The complete  1loop EW contributions to the real parts of the
  four HC amplitudes listed in
(\ref{4HC-amp-list}), and their  supersimple (sim) approximations,
in  the MSSM benchmark described in the text; (as shown in  Fig.\ref{HC-full-amp},
the  SM and MSSM results are very close to each other).
The new physics AGC1, AGC2 and $Z'$ results are also presented.
Upper (lower) panels describe the TT (LL)
amplitudes respectively. Imaginary parts of the amplitudes are much smaller and they
are not shown. }
\label{HC-full-NPamp}
\end{figure}

\end{document}